\documentclass{article}

\usepackage{fullpage}

\usepackage{fullpage}
\usepackage{times}
\usepackage{soul}
\usepackage{url}
\usepackage[hidelinks]{hyperref}
\usepackage[utf8]{inputenc}
\usepackage[small]{caption}
\usepackage{graphicx}
\usepackage{amsfonts}
\usepackage{amsmath}
\usepackage{amsthm}
\usepackage{booktabs}

\usepackage[numbers, sort]{natbib}
\usepackage[switch]{lineno}

\usepackage[ruled,linesnumbered]{algorithm2e} %
\usepackage{booktabs}
\usepackage{multirow}
\usepackage{algpseudocode}
\usepackage{amsmath}
\usepackage{bbm}
\usepackage{epsfig}
\usepackage{hyperref}
\usepackage{adjustbox}
\usepackage{subcaption}
\usepackage{url}
\usepackage{xspace}
\usepackage{tikz}
\usepackage[framemethod=TikZ]{mdframed}
\usepackage{balance}
\usetikzlibrary{bayesnet}
\usetikzlibrary{positioning}
\usepackage{pifont}%
\newcommand{\cmark}{\ding{52}}%
\newcommand{\xmark}{\ding{55}}%

\hypersetup{
    unicode=false,          %
    pdftoolbar=true,        %
    pdfmenubar=true,        %
    pdffitwindow=false,     %
    pdfstartview={FitH},    %
    pdftitle={My title},    %
    pdfauthor={Author},     %
    pdfsubject={Subject},   %
    pdfcreator={Creator},   %
    pdfproducer={Producer}, %
    pdfnewwindow=true,      %
    colorlinks=true,       %
    linkcolor=black,          %
    citecolor=black,        %
    filecolor=black,      %
    urlcolor=black           %
}

\newcommand{\spara}[1]{\smallskip\noindent{\bf #1}}

\newtheorem{definition}{Definition}
\newtheorem{proposition}{Proposition}

\newtheorem{theorem}{Theorem}
\newtheorem{corollary}{Corollary}

\newtheorem{lemma}{Lemma}
\newtheorem{problem}{Problem}

\renewcommand{\vec}[1]{\ensuremath{\mathbf{#1}}\xspace}
\newcommand{\mat}[1]{\ensuremath{\mathbf{#1}}\xspace}
\newcommand{\eye}{\mat{I}}

\newcommand{\Din}{\ensuremath{\mat{D}^{\mathrm{in}}}\xspace}
\newcommand{\Dout}{\ensuremath{\mat{D}^{\mathrm{out}}}\xspace}

\newcommand{\convexset}{\ensuremath{\mathcal{C}}\xspace}

\newcommand{\squishlist}{
 \begin{list}{$\bullet$}
  {  \setlength{\itemsep}{0pt}
     \setlength{\parsep}{3pt}
     \setlength{\topsep}{3pt}
     \setlength{\partopsep}{0pt}
     \setlength{\leftmargin}{2em}
     \setlength{\labelwidth}{1.5em}
     \setlength{\labelsep}{0.5em}
} }
\newcommand{\squishlisttight}{
 \begin{list}{$\bullet$}
  { \setlength{\itemsep}{0pt}
    \setlength{\parsep}{0pt}
    \setlength{\topsep}{0pt}
    \setlength{\partopsep}{0pt}
    \setlength{\leftmargin}{2em}
    \setlength{\labelwidth}{1.5em}
    \setlength{\labelsep}{0.5em}
} }

\newcommand{\squishdesc}{
 \begin{list}{}
  {  \setlength{\itemsep}{0pt}
     \setlength{\parsep}{3pt}
     \setlength{\topsep}{3pt}
     \setlength{\partopsep}{0pt}
     \setlength{\leftmargin}{1em}
     \setlength{\labelwidth}{1.5em}
     \setlength{\labelsep}{0.5em}
} }

\newcommand{\squishend}{
  \end{list}
}
\usepackage[many]{tcolorbox}
\usepackage{xcolor}

\newcommand{\compilehidecomments}{true}%

\ifthenelse{ \equal{\compilehidecomments}{false} }{%
	\newcommand{\francesco}[1]{}	
        \newcommand{\wei}[1]{}
        \newcommand{\federico}[1]{}
	\newcommand{\yuko}[1]{}
	\newcommand{\atsushi}[1]{}
}{

	\newcommand{\francesco}[1]{{\color{orange} [\text{Francesco:} #1]}}
	\newcommand{\federico}[1]{{\color{blue}  [\text{Federico:} #1]}}
	\newcommand{\yuko}[1]{{\color{violet} [\text{Yuko:} #1]}}
	\newcommand{\atsushi}[1]{{\color{teal} [\text{Atsushi:} #1]}}
}

\title{Minimizing Polarization and Disagreement\\ in the Friedkin--Johnsen Model with Unknown Innate Opinions}

\date{}

\author{
Federico Cinus \\
Sapienza University, Rome, Italy \\
CENTAI Institute, Turin, Italy \\
\texttt{cinus@diag.uniroma1.it}
\and
Atsushi Miyauchi \\
CENTAI Institute, Turin, Italy \\
\texttt{atsushi.miyauchi@centai.eu}
\and
Yuko Kuroki\\
CENTAI Institute, Turin, Italy \\
\texttt{yuko.kuroki@centai.eu}
\and
Francesco Bonchi \\
CENTAI Institute, Turin, Italy \\
\texttt{bonchi@centai.eu}
}

\begin{document}

\maketitle
\begin{abstract}
The bulk of the literature on opinion optimization in social networks adopts the Friedkin–Johnsen (FJ) opinion dynamics model, in which the innate opinions of all nodes are known: this is an unrealistic assumption.
In this paper, we study opinion optimization under the FJ model without the full knowledge of innate opinions. Specifically, we borrow from the literature a series of objective functions, aimed at minimizing polarization and/or disagreement, and we tackle the budgeted optimization problem, where we can query the innate opinions of only a limited number of nodes.
Given the complexity of our problem, we propose a framework based on three steps: (1) select the limited number of nodes we query, (2) reconstruct the innate opinions of all nodes based on those queried, and (3) optimize the objective function with the reconstructed opinions. For each step of the framework, we present and systematically evaluate several effective strategies. A key contribution of our work is a rigorous error propagation analysis that quantifies how reconstruction errors in innate opinions impact the quality of the final solutions.
Our experiments on various synthetic and real-world datasets show that we can effectively minimize polarization and disagreement even if we have quite limited information about innate opinions.
\end{abstract}

\section{Introduction}
\label{sec:intro}
The synergetic effect of natural \emph{homophily} and the algorithms employed by social media platforms, e.g., \emph{who-to-follow} recommender systems and \emph{``feed''} content rankers, largely constitutes the information diet of social media users, aligning it with their own opinions.
This, together with the natural tendency to \emph{confirmation bias} \cite{del2017modeling},
leads to the so-called \emph{``echo-chamber''} effect~\citep{quattrociocchi2016echo,cinus2022effect},
where individuals with similar mindsets reciprocally reinforce their pre-existing beliefs,  which in turn leads to \emph{polarization}~\cite{nikolov2015measuring,pariser2011filter}. The rising awareness of the societal risks of extreme polarization driven by social media has spurred a great deal of research on algorithmic interventions aimed at mitigating these harmful effects  \cite{hartman2022interventions,GarimellaMGM18,AslayMGG18,garimella2017balancing,TuAG20}.
While the bulk of this literature focuses on a static
setting, a growing body of work takes into consideration the dynamic nature of the underlying opinion-formation process \cite{musco2018minimizing,cinus2023rebalancing,chen2018quantifying,zhu2021minimizing,xu2021fast}, in particular adopting the widely used \emph{Friedkin--Johnsen (FJ)} opinion-dynamics model~\cite{friedkin1990social}.

\spara{Background and Related Work.}
In the FJ model, social media users are depicted as nodes in a network and social ties are represented by edges. Each individual has an \emph{innate
opinion}, which may differ from their \emph{expressed opinion} on
social media, due to various factors, such as social pressure or fear
of judgement. The model operates through an iterative process where users adjust their expressed opinions by taking a weighted average of their own innate opinion and the expressed opinions of their connected peers. It is well known that the equilibrium state of expressed opinions has an analytic form based on the Laplacian of the network and the innate opinions~\cite{musco2018minimizing}. 
Due to its linear algebraic nature, the model has inspired several optimization problems involving susceptibility~\cite{abebe2021opinion,marumo2021projected}, stubbornness~\cite{xu2022effects}, exposure timelines~\cite{zhou2024modeling}, and adversarial attacks~\cite{tu2023adversaries}.
Additionally, the model has led to some generalizations such as randomized interactions~\cite{fotakis2016opinion}, dynamic social pressure~\cite{ferraioli2017social}, and discrete opinion settings~\cite{chierichetti2013discrete,auletta2016generalized}.

In the literature on optimization problems under the FJ model, the seminal work of \citet{musco2018minimizing} introduced the problem of minimizing the sum of \emph{polarization} and \emph{disagreement} by intervening the weights of edges, and designed a polynomial-time approximation scheme based on the convexity of the objective function. Specifically, they investigated two problems: the first problem requires to find a weighted undirected graph that optimizes the objective function, given a total edge-weight budget and without considering a specific input network, while the second aims to optimize users' innate opinions.  Following up on this work, several modeling and intervention strategies have been explored~\cite{chen2018quantifying,abebe2021opinion,zhu2021minimizing,cinus2023rebalancing,zhou2024modeling,xu2021fast}.

Most relevant to our work, \citet{cinus2023rebalancing} recently extended the problem of \citet{musco2018minimizing}  to general directed networks, where the intervention is on the weights of out-going edges of each node, i.e., \emph{rebalancing} the relative importance of the accounts that the user follows, so as to calibrate the frequency with which the contents produced by various accounts are shown in the \emph{social feed} of the user. %

\spara{Our Contributions.}
All of this body of work on opinion optimization under the FJ model \emph{assumes that the innate opinions of the nodes are all known and given as input}. 
However, in practical scenarios, obtaining such information is a task inherently imprecise and expensive. For example, analyzing user opinions on a controversial topic (e.g., COVID-19 vaccination or Brexit) on a social media platform would require either large-scale surveys or extensive behavioral analysis (e.g., posts, reposts, and likes on platforms like $\mathbb{X}$). 
Furthermore, even in scenarios in which one is able to reconstruct all the opinions, how the inherent error in such opinion reconstruction influences the performance in the opinion optimization task, has not been addressed in the literature.

In this paper, to fill this gap, we consider a series of opinion optimization problems \emph{without the full knowledge of innate opinions},
and tackling the budgeted optimization problems, where we can query the innate opinions of only a limited number of nodes.
As objective functions to be minimized, following~\cite{chen2018quantifying,musco2018minimizing,zhu2021minimizing,cinus2023rebalancing}, we consider polarization, disagreement, and the sum of the two in both directed and undirected networks, for a total of six different objectives. As an intervention mechanism,  following \cite{cinus2023rebalancing}, we consider re-weighting the relative importance of the accounts that each user follows.

A crucial step towards our solution is to effectively reconstruct the innate opinions of nodes that we did not query. In this regard, it is worth mentioning the very recent work of \citet{neumann2024sublinear}, which studies opinion estimation in the FJ model, without considering opinion optimization.
Their estimation approach is not directly applicable to our set of opinion optimization problems, which require gradient descent methods as in \cite{musco2018minimizing,cinus2023rebalancing}:
these methods require
numerous evaluations of both the objective values and the gradients of the objective function across different solution,
resulting in significant computational demands and even requiring a non-trivial adaptation of their algorithm.
Finally, \citet{neumann2024sublinear} assume to have access to an oracle for the expressed opinions (which we do not have) and consider only undirected networks, while we study the optimization problems also on directed graphs.

\spara{Roadmap.} In  \S\ref{sec:problem}, we provide the necessary background on the FJ model and introduce the six objective functions we consider and the formal statement of our problem.

In \S\ref{sec:characterization}, we propose a pipeline that integrates innate opinion reconstruction with opinion optimization and theoretically evaluate its impact on final solutions. We characterize the objectives in terms of convexity and Lipschitz continuity, deriving their solvability and an upper bound on optimization error based on opinion reconstruction error.

In \S\ref{sec:methods},  we present methods for selecting nodes to query their innate opinions and reconstructing the opinions of unqueried nodes. For node selection, we use heuristics based on centrality measures. For opinion reconstruction, motivated by the strong \emph{homophily} of innate opinions in real-world networks, we apply strategies including label propagation~\cite{zhu2002learning}, graph neural networks~\cite{kipf2016semi}, and graph signal processing~\cite{lorenzo2018sampling}.

Finally, in \S\ref{sec:experiments}, we perform extensive computational experiments on synthetic and real-world datasets with up to 1.6 million edges. Our key finding is that opinions can be effectively optimized even with limited information about innate opinions, using appropriate combinations of node selection strategies and opinion reconstruction methods.

\section{Problem Definition}
\label{sec:problem}
In this section, we first revisit the \emph{Friedkin--Johnsen (FJ) model}~\cite{friedkin1990social}, followed by six objective functions from the literature, i.e., polarization, disagreement, and their combination, 
in directed networks \cite{cinus2023rebalancing} and undirected networks~\cite{musco2018minimizing,chen2018quantifying,xu2021fast}. 
Finally, we define the problem to be addressed in this paper.

Although the FJ model is typically presented in the undirected case, we here focus on the more general and interesting case of directed graphs, following the treatment of \citet{cinus2023rebalancing}. We thus consider an edge-weighted directed graph $G=(V,E)$, with $n=|V|$ nodes and $m=|E|$ edges, where each node $i\in V$ corresponds to a user, and
each directed edge $(i,j)\in E$ indicates
that $i$ \emph{``follows''} $j$ or, in other words, that  $j$ can influence the opinion of $i$. For $(i,j)\in E$,
the edge weight $a_{ij}$
quantifies how strongly user $j$ influences user $i$.
We assume that $a_{ij} > 0$ if $(i,j)\in E$ and $a_{ij} = 0$ if $(i,j)\notin E$
and we represent all the weights as a matrix $\mat{A}$, i.e., $\mat{A}[i,j] = a_{ij}$.
In the FJ model, each node $i\in V$ has an \emph{innate opinion} $s_i$ about one topic, which
may differ from their \emph{expressed opinion} $z_i$ on social media about the same topic,
due to various factors, such as social pressure or fear of judgement.
The sets of innate and expressed opinions, for all nodes in the network,
are represented by vectors $\vec{s}\in\mathbb{R}^{{n}}$ and $\vec{z}\in\mathbb{R}^{{n}}$, respectively. The nodes update their expressed opinions,
based on the expressed opinions of their neighbors and their own innate opinions.
Specifically, for each node $i\in V$, its expressed opinion $z_i$ at time $t+1$
is given by the average of the expressed opinions of its neighbors at time $t$
and its own innate opinion, weighted by the strength of their influence.
If we denote by \Dout the diagonal matrix whose $i$-th diagonal entry
is the weighted out-degree of node~$i$, i.e.,
$\Dout[i,i] = \sum_{j\in V} \mat{A}[i,j]$,
and by $\vec{z}^{(t)}$ the vector of the expressed opinions at time $t$,
the opinion-update rule can be written in matrix notation as
\begin{equation}
\vec{z}^{(t+1)} = (\Dout+\eye)^{-1}(\mat{A}\vec{z}^{(t)} + \vec{s}).
\label{eq:update_matrix}
\end{equation}

By iterating Eq.~\eqref{eq:update_matrix} and using the matrix convergence theorems~\cite[Theorem 7.17 and Lemma 7.18]{burden2015numerical}, 
we can find the equilibrium of the system, where the opinions of all nodes have converged to a steady state. Specifically, the equilibrium is given by
\begin{equation}
\vec{z}^* = (\eye + \mat{L})^{-1}\vec{s},
\label{eq:equilibrium}
\end{equation}
where $\mat{L}=\Dout-\mat{A}$ is the Laplacian matrix of $G$~\cite{cinus2023rebalancing}.
Eq.~\eqref{eq:equilibrium} shows that the equilibrium opinions depend only on the innate opinions
and the structure of the social network.

Following the literature \cite{cinus2023rebalancing}, we assume that the adjacency matrix $\mat{A}$ of the directed graph $G$ is row-stochastic, i.e., $\mat{A}\vec{1} = \vec{1}$, where $\vec{1}$ denotes the all-ones vector. This assumption allows for a straightforward interpretation that the total amount of influence each node receives sums to 1.
In this case, the Laplacian $\mat{L}$ is given by $\mat{L}=\mat{D}^\mathrm{out}-\mat{A}=\mat{I}-\mat{A}$. Thus, the equilibrium opinion is written as
\begin{equation*}
\vec{z}^* = (2\mat{I} - \mat{A})^{-1} \vec{s}.
\end{equation*}
Finally, as in the literature~\cite{musco2018minimizing,cinus2023rebalancing}, we assume that all opinions are mean-centered, i.e., $\sum_{u\in V}z^*_u=0$.

We are now ready to introduce the six objective functions.

\begin{definition}[Polarization for directed graphs (\textsc{P-Dir})] The polarization for the equilibrium opinion vector $\vec{z}^*$ is defined to be the deviation of the opinions of nodes from the average opinion~\cite{musco2018minimizing}.
As the opinions are mean-centered, the polarization at the equilibrium is defined as $\sum_{u\in V}{z^*_u}^2={\vec{z^*}}^{\top}{\vec{z^*}}$. Following \cite{cinus2023rebalancing}, the objective function of P-Dir is given by
\begin{equation}
    f(\vec{s},  \mat{L}\!=\!\mat{I}\!-\!\mat{A}) := \vec{s}^\top (2\mat{I} - \mat{A})^{-\top} (2\mat{I} - \mat{A})^{-1} \vec{s}. 
    \label{eq:p-dir}
\end{equation}
\end{definition}

\begin{definition}[Disagreement for directed graphs (\textsc{D-Dir})]  Following \cite{cinus2023rebalancing}, the disagreement for the equilibrium opinion vector $\vec{z^*}$ is defined to be the degree of difference of neighboring opinions on $G$ given by $\sum_{(u,v)\in E}a_{uv}(z^*_u-z^*_v)^2=\vec{z^*}^{\top}(\mat{I}+\mat{D}^\mathrm{in}-2\mat{A}) \vec{z^*}$, where $\mat{D}^\mathrm{in}$ is the in-degree counterpart of $\mat{D}^\mathrm{out}$.  Then, the objective function of D-Dir is
\begin{equation}
 \! \!  f(\vec{s}, \mat{L}\!=\!\mat{I}\!-\!\mat{A})\! := \! \frac{1}{2}\vec{s}^{\top} (2\mat{I}\!-\!\mat{A})^{-\top} \!  (\mat{I}\!+\!\Din\!-\!2\mat{A}) (2\mat{I}\!-\!\mat{A})^{-1} \vec{s}. 
    \label{eq:d-dir}
\end{equation}
\end{definition}
\begin{definition}[Polarization plus Disagreement for directed graphs (\textsc{PD-Dir})] 
We define polarization plus disagreement as the sum of polarization~\eqref{eq:p-dir} and disagreement~\eqref{eq:d-dir}, as in~\cite{cinus2023rebalancing}: 
\begin{equation}
   f(\vec{s},  \mat{L}\!=\!\mat{I}\!-\!\mat{A})\!:=\!\frac{1}{2} \vec{s}^{\top} (2\mat{I} - \mat{A})^{-\top} (\eye - \Din ) (2\mat{I}\!-\!\mat{A})^{-1} \vec{s}.
   \label{eq:pd-dir}
\end{equation}
\end{definition}

\spara{The Undirected Case.} Next, we consider the undirected version of the three objective functions above. 
Undirected graphs are useful to model social networks in which each link represents a bidirectional ``friendship'' relation. 
The formalization used so far applies straightforwardly to the undirected case by considering any undirected edge $\{u,v\}$ as the two directed edges $(u,v)$ and $(v,u)$. 
In the undirected case, we can use the symmetry of $\mat{A}$ and $\mat{L}$ to simplify the notation. 
Following \citet{musco2018minimizing}, the objective functions for undirected graphs are then defined as follows: 

\begin{definition}[Polarization for undirected graphs (\textsc{P-Undir})]
\begin{equation}
    f(\vec{s},  \mat{L}) := \mathbf{s}^{\top} (\mat{I} + \mathbf{L})^{-2}\mathbf{s}. 
    \label{eq:p-und}
\end{equation}
\end{definition}

\begin{definition}[Disagreement for undirected graphs (\textsc{D-Undir})]
\begin{equation}
    f(\vec{s},  \mat{L}) := \mathbf{s}^{\top} (\mat{I} + \mathbf{L})^{-1} \mat{L}(\mat{I} + \mathbf{L})^{-1}\mathbf{s}. 
    \label{eq:d-und}
\end{equation}
\end{definition}

\begin{definition}[Polarization plus Disagreement for undirected graphs (\textsc{PD-Undir})]
\begin{equation}
    f(\vec{s},  \mat{L}) := \mathbf{s}^\top (\mat{I} +\mathbf{L})^{-1}\mathbf{s}. 
    \label{eq:pd-und}
\end{equation}
\end{definition}

\subsection{Our Problem}
Given a graph $G$ and influence weights \mat{A} along its edges,
our goal is to adjust the edge weights \mat{A} so as to minimize
one of the six objective functions above. As discussed in Introduction, our intervention corresponds to re-weighting the relative importance of the accounts that each user follows, so as to calibrate the frequency with which the contents produced by various accounts are shown to the user. Earlier works have studied similar tasks
on undirected~\cite{chen2018quantifying,musco2018minimizing} and directed graphs \cite{cinus2023rebalancing}. However, the problem we consider is much more complex, as we assume that \emph{we have no prior knowledge of the innate opinions $\vec{s}$}, and instead, we are given a \emph{budget} $b\in \mathbb{Z}_{>0}$ that represents the number of nodes we can query their innate opinions.
We assume that if we query the innate opinion $s_v$ for $v\in V$, we can obtain the exact value of $s_v$.

Following \citet{cinus2023rebalancing}, 
we restrict the feasible set of solutions to adjacency matrices where
the set of edges is a subset of the edges in the input graph and 
the out-degree of each node is preserved.
By doing so, we are asked to 
use only pre-existing links and
preserve the total engagement of each user in the social network.
Formally, given the adjacency matrix \mat{A} for a directed graph $G$,
we define the convex set of feasible solutions as follows: 
\begin{equation*}
\convexset(\mat{A}) =
\{ \mat{X} \in \mathbb{R}_{\geq 0}^{{n}\times {n}} \mid
\mat{X}\vec{1}=\mat{A}\vec{1},\: \mat{A}[i,j]=0 \implies \mat{X}[i,j]=0 \}.
\end{equation*}

\noindent In the case of undirected graphs, the feasible set is 
\begin{equation*}
\convexset(\mat{L}) =
\{ \mat{X} \in \mathcal{L}^n \mid
\mathrm{Tr}(\mat{X})=\mathrm{Tr}(\mat{L}), \:
\mat{L}[i,j]=0 \implies \mat{X}[i,j]=0 \}, 
\end{equation*}

\noindent 
\hspace{-1.2mm}where $\mathcal{L}^n$ is the set of Laplacians for graphs with $n$ nodes. Here we adopt the approach of ~\citet{musco2018minimizing}, where the weighted degree of the nodes is not necessarily preserved.

We are now ready to define the problem. 
\begin{problem}[Polarization and Disagreement Minimization with Unknown Innate Opinions]
We are given an edge-weighted directed (resp. undirected) graph $G = (V,E)$ with the unknown innate opinion vector $\vec{s}\in \mathbb{R}^n$.
We are also given a budget $b\in \mathbb{Z}_{>0}$.
The goal is to find a new adjacency matrix $\mat{A}^*\in \convexset(\mat{A})$ (resp. Laplacian $\mat{L}^*\in \convexset(\mat{L})$) that minimizes the specified objective function among Eqs.~\eqref{eq:p-dir}--\eqref{eq:pd-dir} (resp. Eps.~\eqref{eq:p-und}--\eqref{eq:pd-und}) defined for the unknown innate opinion vector $\vec{s}$,
under the assumption that we can query the innate opinions of at most $b$ nodes.
\end{problem}

\section{Characterization}
\label{sec:characterization}
Given the complexity of our problem, we propose a framework based on three steps (which we will detail in Section~\ref{sec:methods}):  (1) select $b$ nodes and observe their innate opinions, (2) reconstruct the innate opinions of all nodes, using the $b$ observed opinions and the network structure, and (3) optimize the objective function with the reconstructed opinions.

In this section, to characterize the importance of steps (1) and (2), we quantify how the reconstruction error of innate opinions affects the final quality of solutions.
The main result of our analysis is presented in Theorem~\ref{theorem:propagation}, revealing 
how the reconstruction error together with the Lipschitz constant of the objective function bound the error of optimization. 
To complete the analysis, we then derive the Lipschitz constant of each objective function with respect to the opinion vectors. %

All proofs are deferred to the Supplementary Material.

\subsection{Hardness}\label{sec:node_sel}
First, we discuss the hardness of the node selection problem for opinion reconstruction and the non-convexity of most objectives. These results build a taxonomy of objective functions related to the FJ model, as summarized in Table~\ref{tab:objectives}.

\spara{Node Selection for Opinion Reconstruction.}
In general, selecting an optimal subset of nodes to query, which minimizes the error of some estimate, is a well-known problem in optimal design~\cite{pukelsheim2006}.
This involves choosing a subset of $b$ sample locations to recover the unknown parameter \(\vec{s}\).
When experiments are selected integrally, as in our problem, standard criteria such as A-optimal design, D-optimal design, and E-optimal design are known to be NP-hard, even in the simplest cases like linear experiments~\cite{pmlr-v99-madan19a}. 
This implies the potential difficulty of node selection in our problem.

\spara{Non-Convexity of Objective Functions.}
We next state the non-convexity of our objectives, except for \textsc{PD-Undir} being convex, related to the solvability of the problems.

\begin{proposition}
\label{prop:nonconvex-obj}
The objectives~\eqref{eq:p-dir}--\eqref{eq:d-und} are not matrix-convex.
\end{proposition}

\begin{proposition}
\label{prop:convex-obj}
The objective~\eqref{eq:pd-und} is matrix-convex.
\end{proposition}

\subsection{Error Analysis}
Let $\vec{s}$ and $\hat{\vec{s}}$ be the true innate opinions and the reconstructed innate opinions, respectively.
We define the \emph{reconstruction error} of $\hat{\vec{s}}$ to be $\|\vec{s}-\hat{\vec{s}}\|$. 
In general, a function $f$ on $\mathbb{R}^n$ is said to be $K$-\emph{Lipschitz continuous} if there exists a constant $K$ such that for all \( \vec{x}, \vec{y} \) in its domain, \( |f(\vec{x}) - f(\vec{y})| \leq K \|\vec{x} - \vec{y}\| \), where the parameter $K$ is called a \emph{Lipschitz constant}. 
The following theorem shows %
how the reconstruction error together with the Lipschitz constant of the objective function bound the error of optimization. 
\begin{theorem}
\label{theorem:propagation}
Let $\mat{L}^*$ and $\hat{\mat{L}}$ be optimal solutions, which minimize $f(\vec{s},\mat{L})$ and $f(\hat{\vec{s}}, \mat{L})$ in Eqs.~\eqref{eq:p-dir}--\eqref{eq:pd-und}, respectively, over the feasible set defined. 
Suppose that $f$ is $K$-Lipschitz continuous with respect to the first argument.
Then, it holds that $f(\vec{s},\hat{\mat{L}})-f(\vec{s},\mat{L}^*)\leq 2 K \|\vec{s}-\hat{\vec{s}}\|$. 
\end{theorem}

\begin{table}[t]
\centering
\caption{Summary of the six objectives functions, convexity w.r.t. the Laplacian matrix $\mathbf{L}$, gradients w.r.t. the opinion vector $\vec{s}$, and their corresponding Lipschitz constants $K$.}
\vspace{-2mm}

\begin{tabular}{llll}
\toprule
Objective & Convex  & Gradient w.r.t. $\vec{s}$  & $K$ \\
\toprule

\textsc{P-Dir} &  \xmark & $2(2\mat{I} - \mat{A})^{-\top} (2\mat{I} - \mat{A})^{-1} \vec{s}$ & 2 \\
\midrule
\textsc{D-Dir} &  \xmark & $(2\mat{I} - \mat{A})^{-1}  (\eye + \Din - 2\mat{A}) (2\mat{I} - \mat{A})^{-1} \vec{s}$ & $1+\Delta(G)$ \\
\midrule
\textsc{PD-Dir} &  \xmark & $(2\mat{I} - \mat{A})^{-T} (\mat{I} - \mat{\Din}) (2\mat{I} - \mat{A})^{-1} \vec{s}$ & 1 \\
\midrule
\textsc{P-Undir} &  \xmark & $2(\mat{I} + \mat{L})^{-2}\vec{s}$ & 2 \\
\midrule
\textsc{D-Undir} & \xmark  & $2(\mat{I} + \mat{L})^{-1}\mat{L}(\mat{I} + \mat{L})^{-1}\vec{s}$ & $2 \Delta(G)$ \\
\midrule
\textsc{PD-Undir} & \cmark &  $2(\mat{I} + \mat{L})^{-1}\vec{s}$ & 2 \\
\bottomrule
\end{tabular}

\label{tab:objectives}
\vspace{-2mm}
\end{table}
\noindent This implies the following multiplicative approximation:
\begin{corollary}\label{corollary:propagation}
Under the assumptions of Theorem~\ref{theorem:propagation} together with $f(\vec{s}, \mat{L}^*)\neq 0$, it holds that 
\begin{align*}
\frac{f(\vec{s}, \hat{\mat{L}})}{f(\vec{s}, \mat{L}^*)} \leq 1 + \frac{2 K \|\vec{s} - \hat{\vec{s}}\|}{f(\vec{s}, \mat{L}^*)}.
\end{align*}
\end{corollary}
\noindent The right-hand-side represents an approximation ratio of $\hat{\mat{L}}$ with respect to the true innate opinions $\vec{s}$. 
From Proposition~\ref{prop:nonconvex-obj}, we know that most of the objectives are non-convex, and for those objectives, no exact algorithm is known in the literature. However, for the objective \textsc{PD-Undir}, we can apply Corollary~\ref{corollary:propagation}, due to the convexity of the objective (Proposition~\ref{prop:convex-obj}).

Finally, we provide Lipschitz constants for each objective function in Eqs.~\eqref{eq:p-dir}--\eqref{eq:pd-und}.
Along with Theorem~\ref{theorem:propagation} and Corollary~\ref{corollary:propagation}, 
this establishes an upper bound on the optimization error associated with the reconstruction error.

\begin{proposition}
\label{prop:Lipschitz-constant}
The objective functions in Eqs.~\eqref{eq:p-dir}--\eqref{eq:pd-und} are Lipschitz continuous on the space \( \mathbb{R}^n \) with the following Lipschitz constants: for \textsc{P-Dir}, \( K = 2 \); for \textsc{D-Dir}, \( K = 1 + \Delta(G) \); for \textsc{PD-Dir}, \( K = 1 \); for \textsc{P-Undir}, \( K = 2 \); for \textsc{D-Undir}, \( K = 2 \Delta(G) \); and for \textsc{PD-Undir}, \( K = 2 \); where \( \Delta(G) \) is the maximum (in-)degree of \( G \) (directed or undirected).
\end{proposition}

The proof proceeds as follows: First, we derive the gradient of the objective function with respect to the opinion vector to express the Lipschitz constant in its infinitesimal form. Next, we relate the Lipschitz constant to the spectral norm of the gradient matrix, which can then be bounded.

\section{Methods}
\label{sec:methods}

Here, we outline the three steps of our proposed framework, with detailed explanations, pseudocodes, and time complexity analyses provided in the Supplementary Material.

\subsection{Node Selection Strategies}%

To mitigate computational bottlenecks, we employ three heuristic approaches based on centrality measures, which select the top-$b$ nodes with respect to the following: \textit{Degree Centrality}, which represents the sum of a node’s in-degree and out-degree; \textit{Closeness Centrality}~\cite{bavelas1950communication}, which is inversely proportional to the total shortest path distances from a node to all others; and \textit{PageRank}~\cite{page1999pagerank} with a damping factor of 0.85.

As a \textit{baseline}, we also include a random uniform strategy for node selection, which selects $b$ nodes uniformly at random without leveraging any structural properties of the network.

\subsection{Opinion Reconstruction Methods}
Based on the innate opinions of the $b$ selected nodes, we aim to reconstruct the innate opinions of the remaining nodes as accurately as possible.
To this end, we consider three types of reconstruction methods:  Label Propagation-based, GNN-based, and Graph Signal Processing-based algorithms.

\label{app:node_rec}

\label{app:pseudo}

\spara{Label Propagation (LP).}
We extended LP~\cite{zhu2002learning} to handle continuous values by initializing all node values to zero, setting selected nodes to their true values, and iteratively updating the remaining nodes to the average of their neighbors over a fixed number of iterations.

\spara{Graph Neural Networks (GNN).}
We use a GCN~\cite{kipf2016semi} to reconstruct unknown opinions by propagating known opinions from selected nodes, initializing node features with these values (or zero if unavailable), and training with an MSE loss.

\spara{Graph Signal Processing (GSP).}
We use GSP to reconstruct opinions, assuming the graph signal \( f: V \rightarrow \mathbb{R} \) (opinions in our case) is bandlimited and can be expressed as a linear combination of a limited number of the Laplacian eigenvectors. 
Using mild assumptions for perfect recovery and noise from~\citet{lorenzo2018sampling}, we apply the best linear unbiased estimator (BLUE)~\cite{winer1971statistical} to infer opinions from sampled nodes.

\subsection{Optimization}
In directed graphs, the objective functions~\eqref{eq:p-dir}--\eqref{eq:pd-dir} 
are non-convex, and no approximation algorithm is known in the
literature. To address these computational challenges, we employ a
constrained gradient-descent approach as in~\citet{cinus2023rebalancing}.
In the case of undirected graphs, we formulated the problem with \textsc{PD-Undir} 
as a semidefinite programming (SDP) with CVX as in~\citet{musco2018minimizing}. 
For the other objectives and constraint, we apply the projection steps in \citet{cinus2023rebalancing}.
We compute a local minimum, using the 
reconstructed opinions as input.

\section{Experimental Evaluation}
\label{sec:experiments}
In this section, we evaluate the framework, its strategies, and the impact of the reconstruction error on solution quality compared to the ground-truth opinions. The code is publicly available\footnote{\url{https://anonymous.4open.science/r/QED-submission-B1FD}}, with details on graph instances, user opinions, and parameter settings in
the Supplementary Material. Additional experiments on opinion distribution variations, runtime analysis, sensitivity studies, and comparisons with random baselines are also included in the Supplementary Material.

\subsection{Setup}
We conduct experiments on 16 networks with up to 1.6 million edges, considering both real opinions and synthetic opinions generated with varying distributions and polarization levels. Dataset statistics are provided in Table~\ref{tab:datasets}.

\begin{table}[t]
\caption{Statistics of our datasets. We plot the distribution of the standardized innate opinions (i.e., average opinion is zero). The first three datasets contain direct follow networks on $\mathbb{X}$ and real opinions. The other networks obtained from KONECT~\cite{kunegis2013konect} are associated with opinions sampled from Gaussian distributions.} \label{tab:datasets}
\let\center\empty
\let\endcenter\relax
\centering
\begin{tabular}{lccc}
\toprule
\multirow{2}{*}{\textbf{Dataset}} & \multicolumn{3}{c}{\textbf{Statistics}} \\
        \cmidrule(lr){2-4}
        & $n$ & $m$ & Opinion Distribution \\
\midrule
                               Referendum &  2,479 &  154,831 & {\includegraphics[width=0.15\textwidth, keepaspectratio]{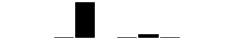}}\\
                                   Brexit &  7,281 &  530,607  & {\includegraphics[width=0.15\textwidth, keepaspectratio]{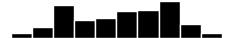}}\\
                                 VaxNoVax & 11,632 & 1,599,220 & {\includegraphics[width=0.15\textwidth, keepaspectratio]{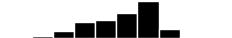}}\\
\midrule
directed/moreno-highschool &    70 &     366    & {\includegraphics[width=0.15\textwidth, keepaspectratio]{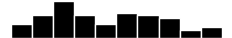}}\\
                directed/wiki-talk-ht &    82 &     154    & {\includegraphics[width=0.15\textwidth, keepaspectratio]{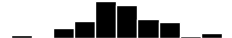}} \\
directed/moreno-innovation &   108 &     510    & {\includegraphics[width=0.15\textwidth, keepaspectratio]{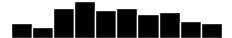}} \\
                directed/moreno-oz &   216 &    2,667   & {\includegraphics[width=0.15\textwidth, keepaspectratio]{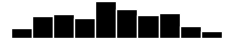}} \\
      directed/librec-filmtrust-trust &   425 &    1,363   & {\includegraphics[width=0.15\textwidth, keepaspectratio]{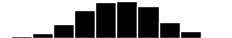}}\\
           directed/dnc-temporalGraph &   949 &    4,029   & {\includegraphics[width=0.15\textwidth, keepaspectratio]{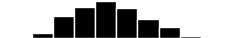}} \\
        directed/librec-ciaodvd-trust &  1,309 &   27,239  & {\includegraphics[width=0.15\textwidth, keepaspectratio]{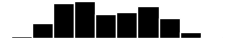}} \\
        directed/moreno-health &  2,298 &   11,999  & {\includegraphics[width=0.15\textwidth, keepaspectratio]{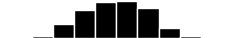}} \\
\midrule
           undirected/ucidata-zachary &    34 &      78 & {\includegraphics[width=0.15\textwidth, keepaspectratio]{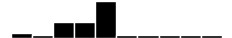}}\\
        undirected/moreno-beach &    43 &     336 & {\includegraphics[width=0.15\textwidth, keepaspectratio]{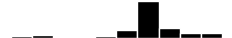}}\\
        undirected/moreno-train &    64 &     243 & {\includegraphics[width=0.15\textwidth, keepaspectratio]{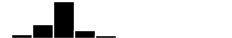}}\\
                       undirected/out.mit &    96 &    2,539 & {\includegraphics[width=0.15\textwidth, keepaspectratio]{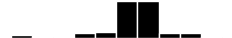}}\\
         undirected/dimacs10-football &   115 &     613 & {\includegraphics[width=0.15\textwidth, keepaspectratio]{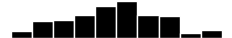}}\\
\bottomrule
\end{tabular}

\end{table}

\spara{Evaluation.}
We evaluate the accuracy of an algorithm as follows. 
Let $\mat{L}_\mathrm{ALG}$ and $\mat{L}^*_\mathrm{ALG}$ be the solutions obtained by the algorithm with the reconstructed innate opinions and the true innate opinions, respectively. 
Note that we can only compute $\mat{L}_\mathrm{ALG}$ when the true innate opinions are unavailable, but $\mat{L}^*_\mathrm{ALG}$ is also computed in experiments for evaluation. 
Then the quality of solution is measured by $\frac{f(\vec{s}, \mat{L}_\mathrm{ALG})}{f(\vec{s}, \mat{L}^*_\mathrm{ALG})}$, 
which we refer to as the \emph{multiplicative error}. 
As the denominator is (even beyond) the best possible achievable by the algorithm, the multiplicative error can be interpreted as a measure of how far the solution's quality deviates from this benchmark. 
Therefore, the smaller the multiplicative error is, the better the solution's quality.

\subsection{Results}

\begin{figure}[t!]
\centering
\includegraphics[width=1.01\columnwidth]{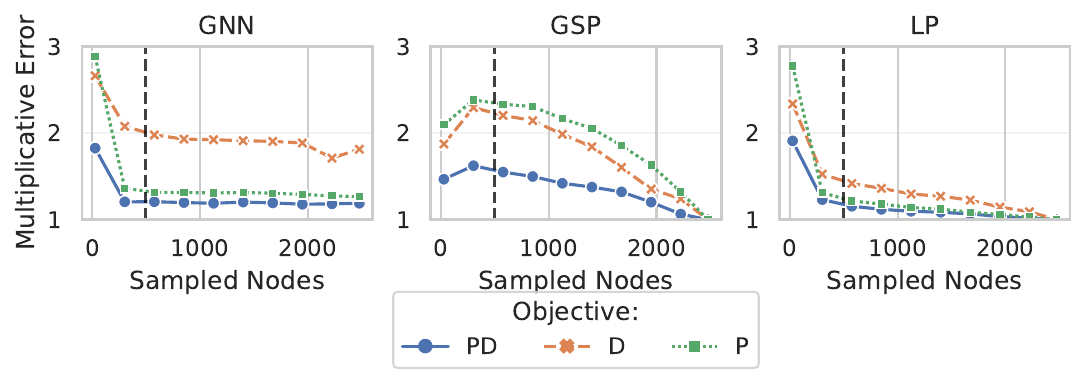}
\caption{Multiplicative error vs. number of sampled nodes in the Referedum dataset.}
\label{fig:sensors-numb}
\end{figure}

\spara{Performance in Real-world Datasets (Directed Graphs).} 
Results are presented in Table~\ref{tab:real-data-recmethods}. We use the three real-world datasets from $\mathbb{X}$, as shown in Table~\ref{tab:datasets},  %
and validate performance across the three objectives for directed graphs in Table~\ref{tab:objectives}. 
Our framework tests three reconstruction methodologies --
LP, GNN, and GSP, from Section~\ref{sec:methods}, using reconstructed opinions from $b = 0.2 |V|$ selected nodes, with this choice validated in Figure~\ref{fig:sensors-numb}. Nodes for opinion reconstruction were selected based on Degree Centrality. Other selections of node sizes and strategies are tested in subsequent experiments.

The results show that LP, despite being the most computationally efficient, consistently achieves the lowest errors across all objectives and datasets, with multiplicative errors ranging from 1.16 to 2.08. 
For the \textsc{P-Dir} objective, LP reduces the error by up to 1.13 compared to GSP, even though this strategy has been optimized for undirected graphs.
For the \textsc{D-Dir} objective, the maximum difference between the multiplicative errors reduces to 0.65, but it is even greater than that in \textsc{PD-Dir} (with a maximum difference of 0.42).
The general trend suggests that the \textsc{D-Dir} objective appears to be the most challenging to optimize, with a maximum multiplicative error exceeding 2. 
This could be due to a dependence between network size and the error bound, as indicated by Proposition~\ref{prop:Lipschitz-constant}, where the function 
$f$ is shown to be Lipschitz continuous with a constant related to $\Delta(G)$.

\begin{table}[t]
\caption{Multiplicative errors for the three objectives (\textsc{D-Dir}, \textsc{P-Dir}, \textsc{PD-Dir}) for 3 real-world directed graphs with different sizes ($n$). 
Opinions are derived from the average stance of tweets users retweeted.} \label{tab:real-data-recmethods}
\let\center\empty
\let\endcenter\relax
\centering
\scalebox{0.75}{
\begin{tabular}{cc|ccc}
\toprule
      & {} & \multicolumn{3}{l}{Multiplicative Error} \\
      & Rec Method &   GNN   &   GSP &    LP           \\
Objective & Network &             &           &        \\
\midrule
\textsc{P-Dir} & Referendum &          \(1.33\) &  \(2.35\) &  \(\mathbf{1.22}\) \\
   & Brexit &             \(2.09\) &  \(2.32\) &  \(\mathbf{1.76}\) \\
   & VaxNoVax &          \(1.85\) &  \(2.36\) &  \(\mathbf{1.41}\) \\
\textsc{D-Dir} & Referendum &        \(1.99\) &  \(2.07\) &  \(\mathbf{1.42}\) \\
   & Brexit &            \(2.66\) &  \(2.53\) &  \(\mathbf{2.08}\) \\
   & VaxNoVax &          \(2.16\) &  \(1.96\) &  \(\mathbf{1.44}\) \\
\textsc{PD-Dir} & Referendum &         \(1.19\) &  \(1.58\) &  \(\mathbf{1.16}\) \\
   & Brexit &             \(1.31\) &  \(1.33\) &  \(\mathbf{1.19}\) \\
   & VaxNoVax &            \(1.33\) &  \(1.39\) &  \(\mathbf{1.16}\) \\
\bottomrule
\end{tabular}
}

\end{table}

\spara{Performance in Semi-Synthetic Datasets (Directed Graphs).}
Results are presented in Table~\ref{tab:directed-opt-gaussian}. We consider the 8 real-world directed networks in Table~\ref{tab:datasets} and a polarized opinion distribution that reflects community structures.

In general, our framework yields multiplicative errors below 2. 
The GNN reconstruction methodology consistently achieves multiplicative errors below this value. 
Nevertheless, LP, while the fastest method, provides the lowest multiplicative errors, except for ``ciaodvd-trust'' and ``wiki talk ht'' networks, where it consistently shows higher multiplicative errors across all three objectives.
\textsc{P-Dir}, at this scale, consistently exhibits the highest errors across all methods and networks. 
The maximum multiplicative errors for the least performing methods in this objective function reach up to 2.74.

Similar trends are observed with uniformly distributed opinions  (Table~\ref{tab:directed-opt-uniform} in the Supplementary Material): LP outperforms the other methods except for ``wiki talk ht'' and ``dnc-temporal'' networks. 

\begin{table}[t]
\caption{Average multiplicative errors for the three objectives (\textsc{D-Dir}, \textsc{P-Dir}, \textsc{PD-Dir}) for 8 real-world directed graphs with different sizes ($|V|$). 
Opinions are Gaussian distributed around a mean corresponding to one of the assigned communities.
This opinions are reconstructed with $b=0.20|V|$ sampled nodes.} 
\label{tab:directed-opt-gaussian}
\let\center\empty
\let\endcenter\relax
\centering
\scalebox{0.75}{
\begin{tabular}{cc|ccc}
\toprule
    &               &  \multicolumn{3}{l}{Multiplicative Error} \\
    &  Rec Method   &        GNN &            GSP &          LP \\
Objective &   Network     &           &                 &              \\
\midrule
\textsc{P-Dir} & highschool      &   \(1.87 \pm 0.40\) &  \(1.97 \pm 0.30\) &  \(\mathbf{1.68 \pm 0.28}\) \\
               & wiki talk ht   &   \(1.28 \pm 0.18\) &  \(\mathbf{1.25 \pm 0.13}\) &  \(1.34 \pm 0.27\) \\
               & innovation     &   \(1.98 \pm 0.22\) &  \(2.06 \pm 0.25\) &  \(\mathbf{1.80 \pm 0.21}\) \\
               & oz  &               \(1.83 \pm 0.17\) &  \(2.74 \pm 0.22\) &  \(\mathbf{1.78 \pm 0.17}\) \\
               & film-trust  &      \(1.28 \pm 0.06\) &  \(1.47 \pm 0.09\) &  \(\mathbf{1.27 \pm 0.06}\) \\
               & dnc-temporal  &     \(1.30 \pm 0.14\) &  \(1.36 \pm 0.08\) &  \(\mathbf{1.25 \pm 0.14}\) \\
               & ciaodvd-trust  &      \(\mathbf{1.56 \pm 0.07}\) &  \(2.50 \pm 0.14\) &  \(2.54 \pm 0.14\) \\
               & health  &             \(1.75 \pm 0.07\) &  \(1.92 \pm 0.07\) &  \(\mathbf{1.57 \pm 0.06}\) \\
   \midrule
\textsc{D-Dir} & highschool   &      \(1.48 \pm 0.20\) &  \(1.48 \pm 0.17\) &  \(\mathbf{1.42 \pm 0.15}\) \\
               & wiki talk ht   &   \(1.24 \pm 0.18\) &  \(\mathbf{1.21 \pm 0.14}\) &  \(1.34 \pm 0.17\) \\
               & innovation   &     \(1.69 \pm 0.17\) &  \(1.71 \pm 0.15\) &  \(\mathbf{1.51 \pm 0.15}\) \\
               & oz   &              \(1.57 \pm 0.15\) &  \(1.93 \pm 0.12\) &  \(\mathbf{1.47 \pm 0.14}\) \\
               & film-trust   &      \(1.23 \pm 0.05\) &  \(1.24 \pm 0.06\) &  \(\mathbf{1.22 \pm 0.05}\) \\
               & dnc-temporal   &     \(1.30 \pm 0.13\) &  \(\mathbf{1.25 \pm 0.11}\) &  \(1.35 \pm 0.11\) \\
               & ciaodvd-trust &     \(\mathbf{1.45 \pm 0.04}\) &  \(1.53 \pm 0.06\) &  \(1.52 \pm 0.03\) \\
               & health  &             \(1.45 \pm 0.06\) &  \(1.50 \pm 0.05\) &  \(\mathbf{1.36 \pm 0.04}\) \\
   \midrule
\textsc{PD-Dir} & highschool    &      \(1.43 \pm 0.15\) &  \(1.42 \pm 0.12\) &  \(\mathbf{1.32 \pm 0.09}\) \\
                & wiki talk ht   &    \(1.23 \pm 0.18\) &  \(\mathbf{1.20 \pm 0.12}\) &  \(1.28 \pm 0.23\) \\
                & innovation   &       \(1.43 \pm 0.11\) &  \(1.39 \pm 0.12\) &  \(\mathbf{1.35 \pm 0.10}\) \\
                & oz   &               \(\mathbf{1.39 \pm 0.09}\) &  \(1.59 \pm 0.09\) &  \(1.40 \pm 0.09\) \\
                & film-trust   &       \(1.16 \pm 0.04\) &  \(1.24 \pm 0.05\) &  \(\mathbf{1.15 \pm 0.03}\) \\
                & dnc-temporal  &       \(1.19 \pm 0.12\) &  \(1.22 \pm 0.08\) &  \(\mathbf{1.16 \pm 0.12}\) \\
                & ciaodvd-trust &       \(\mathbf{1.19 \pm 0.05}\) &  \(1.46 \pm 0.07\) &  \(1.56 \pm 0.05\) \\
                & health  &              \(1.37 \pm 0.02\) &  \(1.37 \pm 0.02\) &  \(\mathbf{1.29 \pm 0.02}\) \\
\bottomrule
\end{tabular}
}

\end{table}

\begin{table}[t]
\caption{Average multiplicative errors in \textsc{PD-Undir} minimization in undirected graphs.} \label{tab:undirected-opt}
\let\center\empty
\let\endcenter\relax
\centering
\begin{tabular}{c|ccc}
\toprule
{}          & \multicolumn{3}{l}{Multiplicative Error (Bound)} \\
Rec Method &                        GNN &              GSP &            LP \\
Network    &                            &                  &               \\
\midrule
zachary       &    \(\mathbf{1.69 \pm 0.29} \:\: (7)\) &      \(2.40 \pm 0.36 \:\: (9)\) &      \(1.70 \pm 0.24 \:\: (8)\) \\
 beach        &    \(\mathbf{2.22 \pm 0.16} \:\: (67)\) &     \(4.03 \pm 0.70 \:\: (94)\) &     \(2.27 \pm 0.20 \:\: (70)\) \\
train         &    \(\mathbf{1.46 \pm 0.13} \:\: (4)\) &      \(2.69 \pm 0.41 \:\: (7)\) &      \(1.82 \pm 0.26 \:\: (6)\) \\
mit           &    \(\mathbf{1.07 \pm 0.11} \:\: (9,540)\) &  \(\mathbf{1.07 \pm 0.11} \:\: (12,067)\) &   \(1.09 \pm 0.15 \:\: (9,071)\) \\
football      &    \(2.11 \pm 0.53 \:\: (3)\) &      \(2.57 \pm 0.72 \:\: (4)\) &      \(\mathbf{1.93 \pm 0.44} \:\: (3)\) \\
\bottomrule
\end{tabular}

\end{table}

\spara{Performance in Semi-Synthetic Datasets (Undirected Graphs).} 
Results are presented in Table~\ref{tab:undirected-opt}. We consider the 5 real-world undirected networks in Table~\ref{tab:datasets} with a polarized opinion distribution that reflects community structures. Network sizes are limited to approximately 100 nodes to avoid the computational bottleneck inherent in the SDP approach for finding an optimal solution.
We test the performance of our framework in minimization of \textsc{PD-Undir}. This problem is well-studied in the literature and includes a standard projection step onto the set of SD matrices~\cite{musco2018minimizing}. For \textsc{P-Undir} and \textsc{D-Undir}, no projection step is known in the literature.

Multiplicative errors show greater variability across different graphs in undirected settings compared to directed ones. This would be because the current optimization problem (\textsc{PD-Undir} minimization) allows the algorithm to find a global optimum, 
resulting in a relatively small value of the denominator in the multiplicative error calculation. 

Bounds on the multiplicative error are presented as averages, following Corollary~\ref{corollary:propagation} and Proposition~\ref{prop:Lipschitz-constant}. 
These bounds are sensitive to the numerical value of the global minimum of the objective. When the minimum is quite small, it can lead to very large bounds that are not practically useful. 
This occurred in 2 out of 5 instances in our experiments, specifically with the ``beach'' and ``mit'' networks.
As a result, the bounds are larger than the actual error, indicating that, in practice, the problem is less challenging than theoretically predicted, and a tighter bound likely exists at this graph size scale.
The bounds on the optimization error are proportional to the reconstruction errors, meaning that better performance is closely linked to improved reconstruction accuracy. For instance, the GNN strategy consistently achieves lower multiplicative errors (up to 1.7 times lower) compared to the GSP method. The LP method is comparable to GNN, except for ``football'' network, where it outperforms the others with a multiplicative error below 2.

\begin{table}[t]
\caption{Multiplicative errors for the three objectives (\textsc{P-Dir}, \textsc{D-Dir}, \textsc{PD-Dir}) for different node selection strategies in real-world datasets.} \label{tab:sampling-strategies-real}
\let\center\empty
\let\endcenter\relax
\centering
\begin{tabular}{cc|cccc}
\toprule
   & {} & \multicolumn{4}{l}{Multiplicative Error} \\
   & Sel Method &  Closeness centrality &    Degree &  PageRank &    Random \\
Objective & Network &                   &           &           &           \\
\midrule
\textsc{P-Dir} & Referendum &         \(1.44\) &  \(1.22\) &  \(\mathbf{1.21}\) &  \(1.27\) \\
               & Brexit &            \(\mathbf{1.74}\) &  \(1.76\) &  \(1.75\) &  \(1.87\) \\
               & VaxNoVax &          \(\mathbf{1.37}\) &  \(1.41\) &  \(\mathbf{1.37}\) &  \(1.61\) \\
\textsc{D-Dir} & Referendum &        \(1.57\) &  \(\mathbf{1.42}\) &  \(1.44\) &  \(1.52\) \\
               & Brexit &            \(2.06\) &  \(2.08\) &  \(\mathbf{2.04}\) &  \(2.35\) \\
               & VaxNoVax &          \(\mathbf{1.41}\) &  \(1.44\) &  \(\mathbf{1.41}\) &  \(1.74\) \\
\textsc{PD-Dir} & Referendum &        \(1.23\) &  \(\mathbf{1.16}\) &  \(1.17\) &  \(1.22\) \\
                & Brexit &            \(\mathbf{1.18}\) &  \(1.19\) &  \(\mathbf{1.18}\) &  \(1.28\) \\
                & VaxNoVax &          \(1.14\) &  \(1.16\) &  \(\mathbf{1.13}\) &  \(1.28\) \\
\bottomrule
\end{tabular}

\end{table}

\spara{Effect of Node Selection Strategies.} 
The results are presented in Table~\ref{tab:sampling-strategies-real}.
We consider real-world datasets to compare different node selection strategies to select $b=0.2|V|$ nodes for reconstruction with the LP method.

On average, PageRank proves to be the most effective strategy for selecting nodes, while Degree Centrality shows consistently strong performance compared to the random strategy.
For \textsc{P-Dir}, using Degree Centrality and PageRank can reduce the multiplicative error by up to 0.2 and 0.24, respectively, compared to random selection. 
For \textsc{D-Dir}, using Degree Centrality can reduce the multiplicative error by up to 0.3 compared to random selection; 0.33 for PageRank. 
For \textsc{PD-Dir}, using Degree Centrality can reduce the multiplicative error by up to 0.12 compared to random selection; 0.15 for PageRank.
Closeness Centrality yields comparable results, although it performs less consistently, particularly on the  ``Referendum'' network.
Results for real directed networks with synthetic opinions are presented in Table~\ref{tab:sampling-strategies-synth-gauss} in the Supplementary Material. These results are consistent, showing the superiority of Degree Centrality and PageRank, except for ``ciaodvd trust" network where the random strategy outperforms the others. 
It is worth noting that selecting nodes uniformly at random tends to cover diverse parts of the network, and given the fact that the innate opinions in real-world networks have a strong homophily, the random strategy would be reasonable and sometimes better than sophisticated ones.
Despite the strong performance of PageRank, Degree Centrality strikes a better balance among quality, consistency, and computational efficiency. As such, we use Degree Centrality as the default strategy in subsequent experiments.

\begin{figure}[t!]
\centering
\begin{minipage}{.502\columnwidth}
  \centering
  \includegraphics[width=\linewidth]{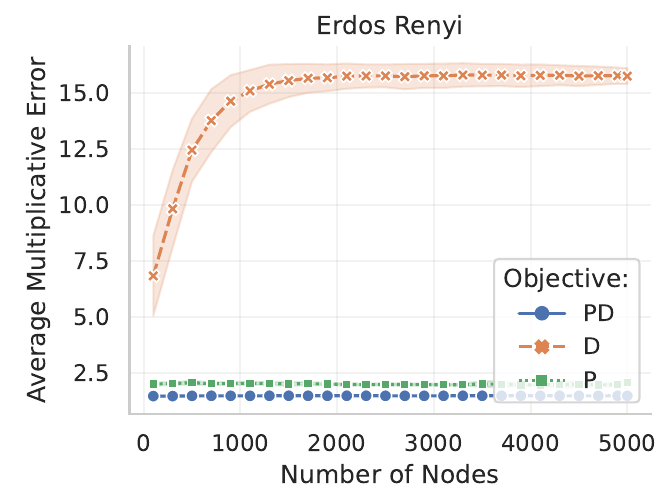}
\end{minipage}%
\hspace{-4pt}
\begin{minipage}{.502\columnwidth}
  \centering
  \includegraphics[width=\linewidth]{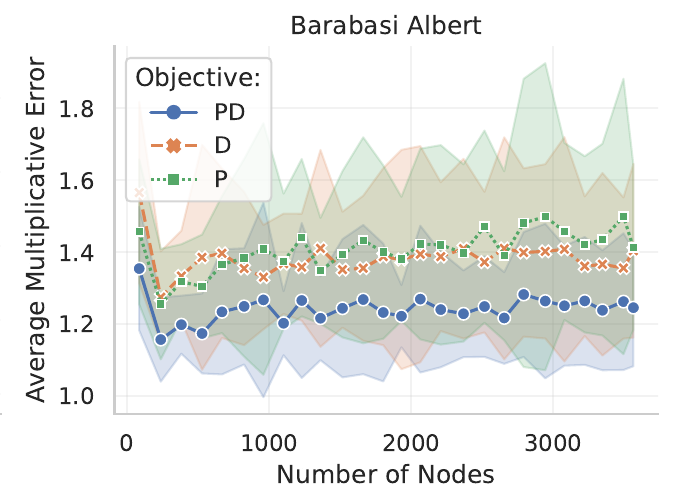}
\end{minipage}
\caption{Average multiplicative errors vs. number of nodes in (left) Erd\H{o}s R\'enyi graph with $p=0.25$, and polarized distribution of opinions; (right) Barab\'asi Albert graph with $m=5$, and polarized distribution of opinions.}
\label{fig:size-dependency}
\end{figure}

\spara{Effect of Network Size.}  
The results are depicted in Figure~\ref{fig:size-dependency}. We consider two synthetic directed networks with sizes ranging from 100 to 5,000 nodes. We create a polarized opinion distribution reflecting community structures and test the capabilities of our framework using the LP strategy. 
We measure the performance in minimizing the three objectives, \textsc{P-Dir}, \textsc{D-Dir}, and \textsc{PD-Dir}, with respect to the number of nodes. A selection strategy based on Degree Centrality is used to select 20\% of nodes in each instance as samples for opinion reconstruction.

Except for \textsc{D-Dir}, multiplicative errors are generally constant, with higher volatility observed in the Barabási-Albert graph. At this range of network sizes, \textsc{D-Dir} shows an increasing error with respect to the number of nodes, followed by a plateau, suggesting that non-constant error is introduced by some network structures.
These observations align with our upper bounds and underscore the importance of characterizing the inter-dependence between network structures and the objectives. This is the first necessary step toward understanding the minimization of such known objectives in an unknown opinion setting, necessitating further analysis and experiments.

\spara{Effect of the number of selected nodes.} 
The results are depicted in Figure~\ref{fig:sensors-numb}. 
We consider the ``Referendum'' dataset to compare different node selection sizes, ranging from 100 up to the size of the network (2,479 nodes).

As expected, the multiplicative error in each objective decreases as the number of selected nodes increases, but different patterns emerge. \textsc{D-Dir} shows the slowest rate of decrease compared to the other objectives across reconstruction strategies. In particular, the GNN strategy exhibits a plateau in the error curve, while LP displays a monotonically decreasing behavior. 
This is why LP has been chosen as the main reconstruction strategy in other experiments, in addition to its computational efficiency. 
The significant drop in multiplicative errors occurs between 15--20\% of the node size. 
The 20\% threshold, indicated in black, represents the selected node size used in all other experiments.
In GSP, the number of selected nodes is linked to the number of frequencies, which is always smaller than the number of selected nodes. A sensitivity analysis is presented in Figure~\ref{fig:sensors-numb-gsignal} in the Supplementary Material.

\section{Conclusions}
\label{sec:conclusions}

This paper contributes to the literature on opinion optimization in social networks under the Friedkin–Johnsen (FJ) model, which assumes innate opinions are fully known. We address the novel problem of opinion optimization under a budget constraint, where the goal is to minimize polarization and/or disagreement by querying a limited number of nodes for their innate opinions. To tackle this, we propose a framework integrating node selection, opinion reconstruction, and optimization, systematically evaluating alternative strategies for each component and identifying the most effective approaches. Our results demonstrate the framework’s practicality, achieving multiplicative errors consistently below 2 and as low as 1.1. Additionally, our error propagation analysis quantifies how reconstruction errors in innate opinions impact the quality of final solutions, offering guidelines for researchers and practitioners, particularly for objectives like disagreement, whose bounds scale with network size.

Although our experiments scale to networks with up to 1.6 million edges, larger real-world networks require more scalable methods. Techniques like GraphSAGE~\cite{hamilton2017inductive} could address this challenge, while active learning strategies adapted to graph structures may improve node selection efficiency. Robustness in heterophilic networks, where dissimilar nodes connect, remains an open challenge, and developing heuristics for such cases is a key avenue for future work. From a theoretical perspective, tighter bounds leveraging network structures that constrain Laplacian eigenvalues could provide stronger guarantees for optimization performance. Finally, practical constraints, such as platform-specific limits on interactions or adjustable edges, require refining the solution space through simple filtering mechanisms to ensure feasibility in real-world applications.

\newpage

\section*{Ethics Statement}

\spara{Societal Impact.}
Our work proposes a framework for opinion optimization under the Friedkin–Johnsen model to mitigate polarization while addressing the challenge of incomplete information. By leveraging the interplay between network structure, innate opinions, and opinion dynamics, this research contributes to efforts aimed at designing interventions that reduce polarization and promote healthier discourse in online environments.

\spara{Ethical Aspects.}
Although this work is grounded in theoretical models and analysis, it carries significant ethical implications. Our experiments are conducted using anonymized datasets to ensure that no personally identifiable information is used or exposed. Nonetheless, methods for inferring opinions inherently carry risks, such as enabling the targeting of individuals based on their inferred opinions. Furthermore, the framework and analysis presented here could, in principle, be repurposed to maximize polarization rather than mitigating it, simply by changing the sign of the objective function.

We acknowledge these risks and emphasize that our work aims to advance ethical and privacy-conscious machine learning. By prioritizing the minimization of polarization and promoting responsible approaches to opinion dynamics, we seek to contribute to societal challenges constructively. Balancing these potential harms, we hope our research sets an example for the development of fairer, more transparent, and ethically grounded interventions in opinion optimization.

\bibliographystyle{named}
\bibliography{reference}

\appendix
\clearpage
\section*{Supplementary Material}
\section{Proofs Omitted from Section 3}

\subsection{Proof of Proposition~\ref{prop:nonconvex-obj}: Convexity Analysis}
We decompose Proposition~\ref{prop:nonconvex-obj} in a series of small lemmas containing specifically created counterexamples to show the non-convexity of the objectives.

We recall that a matrix-valued function $f$ is said to be matrix-convex if and only if it satisfies the following inequality for all $\lambda \in [0, 1]$ and matrices $\mat{A}_1$ and $\mat{A}_2$:
\begin{equation}
    f(\lambda \mat{A}_1 + (1-\lambda) \mat{A}_2) \leq \lambda f(\mat{A}_1) + (1-\lambda) f(\mat{A}_2)
    \label{eq:convex-def}
\end{equation}

\begin{lemma}[P-directed]
The objective function of  Eq.~\eqref{eq:p-dir} is not a matrix-convex function.
\end{lemma}

\begin{proof}
Consider a vector of opinions $\vec{s}^{\top} = (0 \: 1 \: 1)$.
Let $\mat{A}_1$ and $\mat{A}_2$ be two adjacency matrices of two connected graphs:
\[
 \mat{A}_1 =
  \begin{bmatrix}
    0 & 1 &0\\
    1/100& 0 &99/100 \\
    1 & 0 & 0
    \end{bmatrix},
\: \: \text{and} \: \:
\mat{A}_2 =
\begin{bmatrix}
    0 & 1 & 0\\
    1/3& 0 &2/3 \\
    0 & 1 & 0
    \end{bmatrix}.
\]
Setting $\lambda=0.5$, we can compute the two terms in the inequality to obtain
$f((\mat{A}_1 + \mat{A}_2)/2)=1.66$ and
$(f(\mat{A}_1) + f(\mat{A}_2))/2=1.65$.
Thus, the inequality in Eq.~\eqref{eq:convex-def} is violated, and the objective function in Eq.~\eqref{eq:p-dir} is not matrix-convex.
\end{proof}

\begin{lemma}[D-directed]
The objective function of  Eq.~\eqref{eq:d-dir} is not a matrix-convex function.
\end{lemma}

\begin{proof}
Consider a vector of opinions $\vec{s}^{\top} = (1 \: 0 \: -1)$.
Let $\mat{A}_1$ and $\mat{A}_2$ be two adjacency matrices of two connected graphs:
\[
 \mat{A}_1 =
  \begin{bmatrix}
    0 & 1 & 0\\
    2/3& 0 &1/3 \\
    1 & 0 & 0
    \end{bmatrix},
\: \: \text{and} \: \:
\mat{A}_2 =
\begin{bmatrix}
    0 & 1 & 0\\
    1/3& 0 &2/3 \\
    0 & 1 & 0
    \end{bmatrix}.
\]
Setting $\lambda=0.5$, we can compute the two terms in the inequality to obtain
$f((\mat{A}_1 + \mat{A}_2)/2)=0.45$ and
$(f(\mat{A}_1) + f(\mat{A}_2))/2=0.43$.
Thus, the inequality in Eq.~\eqref{eq:convex-def} is violated, and the objective function in Eq.~\eqref{eq:d-dir} is not matrix-convex.
\end{proof}

\begin{lemma}[PD-directed]
The objective function of  Eq.~\eqref{eq:pd-dir} is not a matrix-convex function.
\end{lemma}

\begin{proof}
Consider a vector of opinions $\vec{s}^{\top} = (1 \: 0 \: -1)$.
Let $\mat{A}_1$ and $\mat{A}_2$ be two adjacency matrices of two connected graphs:
\[
 \mat{A}_1 =
  \begin{bmatrix}
    0 & 1 &0\\
    1/100& 0 &99/100 \\
    1 & 0 & 0
    \end{bmatrix},
\: \: \text{and} \: \:
\mat{A}_2 =
\begin{bmatrix}
    0 & 1 & 0\\
    1/3& 0 &2/3 \\
    0 & 1 & 0
    \end{bmatrix}.
\]
Setting $\lambda=0.5$, we can compute the two terms in the inequality to obtain
$f((\mat{A}_1 + \mat{A}_2)/2)=0.82$ and
$(f(\mat{A}_1) + f(\mat{A}_2))/2=0.80$.
Thus, the inequality in Eq.~\eqref{eq:convex-def} is violated, and the objective function in Eq.~\eqref{eq:pd-dir} is not matrix-convex.
\end{proof}

\begin{lemma}[P-undirected]
The objective function of  Eq.~\eqref{eq:p-und} is not a matrix-convex function.
\end{lemma}

\begin{proof}
Consider a vector of opinions $\vec{s}^{\top} = (0 \: 1 \: 1)$.
Let $\mat{A}_1$ and $\mat{A}_2$ be two adjacency matrices of two connected graphs:
\[
 \mat{A}_1 =
  \begin{bmatrix}
    0 & 1/1000 &0\\
    1/1000& 0 &1 \\
    1 & 0 & 0
    \end{bmatrix},
\: \: \text{and} \: \:
\mat{A}_2 =
\begin{bmatrix}
    0 & 1 & 0\\
    1& 0 &1 \\
    0 & 1 & 0
    \end{bmatrix}.
\]
Setting $\lambda=0.5$, we can compute the two terms in the inequality to obtain
$f((\mat{A}_1 + \mat{A}_2)/2)=1.25$ and
$(f(\mat{A}_1) + f(\mat{A}_2))/2=1.11$.
Thus, the inequality in Eq.~\eqref{eq:convex-def} is violated, and the objective function in Eq.~\eqref{eq:p-und} is not matrix-convex.
\end{proof}

\begin{lemma}[D-undirected]
The objective function of  Eq.~\eqref{eq:d-und} is not a matrix-convex function.
\end{lemma}

\begin{proof}
Consider a vector of opinions $\vec{s}^{\top} = (1 \: 0 \: -1)$.
Let $\mat{A}_1$ and $\mat{A}_2$ be two adjacency matrices of two connected graphs:
\[
 \mat{A}_1 =
  \begin{bmatrix}
    0 & 1& 0\\
    1 & 0 & 1 \\
    1 & 0 & 0
    \end{bmatrix},
\: \: \text{and} \: \:
\mat{A}_2 =
\begin{bmatrix}
    0 & 1 & 0\\
    1 & 0 & 1 \\
    0 & 1 & 0
    \end{bmatrix}.
\]
Setting $\lambda=0.5$, we can compute the two terms in the inequality to obtain
$f((\mat{A}_1 + \mat{A}_2)/2)=0.5$ and
$(f(\mat{A}_1) + f(\mat{A}_2))/2=0.49$.
Thus, the inequality in Eq.~\eqref{eq:convex-def} is violated, and the objective function in Eq.~\eqref{eq:d-und} is not matrix-convex.
\end{proof}

\subsection{Proof of Proposition~\ref{prop:convex-obj}: Convexity Analysis}
See the proof of Theorem 1 in ~\citet{musco2018minimizing}.

\subsection{Proof of Theorem~\ref{theorem:propagation}}
\begin{proof}
As $\hat{\mat{L}}$ is a minimizer of $f(\hat{\vec{s}}, \mat{L})$, we have $f(\hat{\vec{s}},\hat{\mat{L}})\leq f(\hat{\vec{s}}, \mat{L}^*)$.
If $f(\vec{s},\hat{\mat{L}})\leq f(\hat{\vec{s}},\hat{\mat{L}})$ holds, we have $f(\vec{s},\hat{\mat{L}})\leq f(\hat{\vec{s}},\mat{L}^*)$,
leading to $f(\vec{s},\hat{\mat{L}})-f(\vec{s},\mat{L}^*)\leq f(\hat{\vec{s}},\mat{L}^*)-f(\vec{s},\mat{L}^*)\leq K\|\vec{s}-\hat{\vec{s}}\|$.
Otherwise,
since $f$ is $K$-Lipschitz continuous, we have
\begin{align*}
f(\vec{s},\hat{\mat{L}})-f(\vec{s},\mat{L}^*)
&\leq |f(\vec{s},\hat{\mat{L}})-f(\hat{\vec{s}},\hat{\mat{L}})| + |f(\hat{\vec{s}},\mat{L}^*)-f(\vec{s},\mat{L}^*)|\\
&\leq 2 K \|\vec{s}-\hat{\vec{s}}\|.
\end{align*}

\end{proof}

\subsection{Proof of Proposition~\ref{prop:Lipschitz-constant}: Lipschitz Constants}
We decompose Proposition~\ref{prop:Lipschitz-constant} in the main text in a series of lemmas: we present their statement and proofs to bound the Lipschitz constants of each objective function studied.
We begin by deriving the gradient of the objective function with respect to the opinion vector to obtain a representation of the Lipschitz constant in its infinitesimal form.

We recall that the Lipschitz constant is associated with the spectral norm of the matrix in the gradient, which corresponds to its maximum eigenvalue.
More formally, a function \( f \) is Lipschitz continuous if there exists a constant \( K \) such that for all \( \vec{s}, \hat{\vec{s}} \) in its domain, \( |f(\vec{s}) - f(\hat{\vec{s}})| \leq K \|\vec{s} - \hat{\vec{s}}\| \).
Equivalently, in its infinitesimal form, the gradient of \( f \) is bounded.
Since we are dealing with quadratic forms, we recall its gradient from Eq. 81 in ~\cite{petersen2008matrix}:
\begin{equation}
    \label{eq:gradient}
    \nabla_{\vec{s}} \vec{s}^{\top} \mat{M} \vec{s} = (\mat{M}+ \mat{M}^{\top})\vec{s}.
\end{equation}

\begin{lemma}[Polarization for directed graphs (P-Dir)]
Given a continuous differentiable function \( f(\vec{s}) = f(\vec{s},  \mat{L}= \eye - \mat{A}) := \vec{s}^\top (2\mat{I} - \mat{A})^{-\top} (2\mat{I} - \mat{A})^{-1} \vec{s} \), where \(\mat{A}\) is a fixed row-stochastic matrix, and \(\vec{s} \in \mathbb{R}^n\),
$f(\vec{s})$ is Lipschitz continuous with $K=2$.
\end{lemma}

\begin{proof}
Using the gradient of a quadratic in Eq.~\eqref{eq:gradient}, the gradient of \( f \) with respect to \( \vec{s} \) is
\begin{align*}
\nabla_{\vec{s}} f(\vec{s}) =
2(2\mat{I} - \mat{A})^{-\top} (2\mat{I} - \mat{A})^{-1} \vec{s}. 
\end{align*}

The product \( (2\mat{I} - \mat{A})^{-\top} (2\mat{I} - \mat{A})^{-1} \) has a spectral norm that is bounded by the product of the spectral norms of the two terms by the submultiplicative property of matrix norms.
Moreover,  eigenvalues of \( (2\mat{I} - \mat{A})^{-1} \) are within \([\frac{1}{3}, 1]\) because the eigenvalues of \( 2\mat{I} - \mat{A} \) are within \([1, 3]\), using the row-stochasticity of $\mat{A}$.
Hence the maximum is $1$.
Thus, the Lipschitz constant \( K \) for \( f \) is bounded by $2$.
\end{proof}

\begin{lemma}[Disagreement for directed graphs (D-Dir) ]
\label{prop:dis-dir}
Given a continuous differentiable function \( f(\vec{s}) = f(\vec{s},  \mat{L}= \eye - \mat{A}) := \frac{1}{2}\vec{s}^{\top} (2\mat{I} - \mat{A})^{-\top}  (\eye + \Din - 2\mat{A}) (2\mat{I} - \mat{A})^{-1} \vec{s} \), where \(\mat{A}\) is a fixed row-stochastic matrix, \(\Din\) is a fixed diagonal matrix with non-negative entries corresponding to the in-degree, and \(\vec{s} \in \mathbb{R}^n\),
$f(\vec{s})$ is Lipschitz continuous with $K=\Delta (G)$.
\end{lemma}

\begin{proof}
Using Eq.~\eqref{eq:gradient}, and the fact that $(\mat{A}\mat{B}\mat{C})^{\top}=\mat{C}^{\top}\mat{B}^{\top}\mat{A}^{\top}$, the gradient of \( f \) with respect to \( \vec{s} \) is
\begin{align*}
\nabla_{\vec{s}} f(\vec{s}) =
(2\mat{I} - \mat{A})^{-1}  (\eye + \Din - 2\mat{A}) (2\mat{I} - \mat{A})^{-1} \vec{s}. 
\end{align*}

The matrices \( (2\mat{I} - \mat{A})^{-\top}, \: (2\mat{I} - \mat{A})^{-1} \) are diagonally-dominant and positive definite, and their spectral norm is bounded by the product of the spectral norms of the two terms by the submultiplicative property of matrix norms.
Moreover,  eigenvalues of \( (2\mat{I} - \mat{A})^{-1} \) are within \([\frac{1}{3}, 1]\) because the eigenvalues of \( 2\mat{I} - \mat{A} \) are within \([1, 3]\), using the row-stochasticity of $\mat{A}$.
Hence the maximum is $1$.
In the other term ($\Din$), the degree matrix is diagonal and hence contains the positive eigenvalues, and we denote its maximum value as $\Delta (G)$.
The adjacency matrix $\mat{A}$ is row-stochastic, hence its minimum eigenvalue is 1, so the negative sign in front of it makes all eigenvalues negative therefore this term does not contribute to the maximum eigenvalue.
Thus, the Lipschitz constant \( K \) for \( f \) is bounded by $1+\Delta (G)$.
\end{proof}

\begin{lemma}[Polarization plus Disagreement for Directed graphs (PD-Dir) ]
Given a continuous differentiable function \( f(\vec{s}) = \frac{1}{2} \vec{s}^T (2\mat{I} - \mat{A})^{-T} (\mat{I} - \Din) (2\mat{I} - \mat{A})^{-1} \vec{s} \), where \(\mat{A}\) is a fixed row-stochastic matrix, \(\Din\) is a fixed diagonal matrix with non-negative entries corresponding to the in-degree, and \(\vec{s} \in \mathbb{R}^n\), $f(\vec{s})$ is Lipschitz continuous with $K=1$.
\end{lemma}

\begin{proof}
Applying the transpose of the product of three matrices, it is easy to see that the matrix inside the quadratic function is symmetric.
Hence, using Eq.~\eqref{eq:gradient}, the gradient of \( f \) with respect to \( \vec{s} \) is \( \nabla_{\vec{s}} f(\vec{s}) = (2\mat{I} - \mat{A})^{-\top} (\mat{I} - \mat{D}) (2\mat{I} - \mat{A})^{-1} \vec{s} \).

The Lipschitz constant \( K \) is associated with the maximum eigenvalue of the matrix in the quadratic form of the gradient.

In the first term ($\eye$), the matrix \( (2\mat{I} - \mat{A})^{-\top} (2\mat{I} - \mat{A})^{-1} \) is symmetric positive definite, and its spectral norm is bounded by the product of the spectral norms of the two terms by the submultiplicative property of matrix norms.  Moreover,  eigenvalues of \( (2\mat{I} - \mat{A})^{-1} \) are within \([\frac{1}{3}, 1]\) because the eigenvalues of \( 2\mat{I} - \mat{A} \) are within \([1, 3]\), using the row-stochasticity of $\mat{A}$. Hence the maximum is $1$.

In the other term ($-\mat{D}$), the degree matrix is diagonal and hence contains the positive eigenvalues. The negative sign makes all eigenvalues negative therefore this term does not contribute to the maximum eigenvalue.

Thus, the Lipschitz constant \( K \) for \( f \) is bounded by 1.
\end{proof}

\begin{lemma}[Polarization for Undirected graphs (P-Und)]
Given a continuous differentiable function \( f(\vec{s}) := \vec{s}^\top (\mat{I} + \mathbf{L})^{-2}\vec{s} \), where \(\mat{A}\) is a fixed row-stochastic matrix and \(\vec{s} \in \mathbb{R}^n\),
the function $f(\vec{s})$
is Lipschitz continuous with $K = 2$.
\end{lemma}

\begin{proof}
Using Eq.~\eqref{eq:gradient}, the gradient of \( f \) with respect to \( \vec{s} \) is
\begin{align*}
\nabla_{\vec{s}} f(\vec{s}) =
2(\mat{I} + \mat{L})^{-2}\vec{s}. 
\end{align*}

The norm of this gradient ($\| 2(\mat{I} + \mat{L})^{-2}\vec{s} \|_2$) is bounded, by sub-multiplicativity, by the eigenvalues of \(  2(\mat{I} + \mat{L})^{-1}(\mat{I} + \mat{L})^{-1} \).
Therefore, by submultiplicativity the Lipschitz constant \( K \) for \( f \) is \( 2 \lambda_{\text{max}}((\mat{I} + \mat{L})^{-1})^2 \).
Using the properties of eigenvalues we know that \(\lambda_{\text{max}}(1+\mat{X})^{-1}=\frac{1}{1+\lambda_{\text{min}}(\mat{X})} \); and since the minimum eigenvalue of a Laplacian matrix is 0, the maximum eigenvalue is 1.
Thus, the Lipschitz constant \( K \) for \( f \) is bounded by $2$.
\end{proof}

\begin{lemma}[Disagreement for undirected graphs (D-Undir)]
Given a continuous differentiable function \( f(\mat{L}, \vec{s}) = \vec{s}^\top (\mat{I} + \mathbf{L})^{-1} \mat{L}(\mat{I} + \mathbf{L})^{-1}\vec{s} \), where \(\mat{L}\) is a fixed symmetric Laplacian matrix and \(\vec{s} \in \mathbb{R}^n\), 
$f(\vec{s})$ is Lipschitz continuous with a Lipschitz constant equal to $2 \Delta(G)$.
\end{lemma}

\begin{proof}
From Eq.~\eqref{eq:gradient}, the gradient of \( f \) is \( \nabla_{\vec{s}} f(\mat{L}, \vec{s}) = 2(\mat{I} + \mat{L})^{-1}\vec{s} \).
The norm of this gradient ($\| 2(\mat{I} + \mat{L})^{-1}\vec{s} \|_2$) is bounded, by the eigenvalues of \(  2(\mat{I} + \mat{L})^{-1}\mat{L}(\mat{I} + \mat{L})^{-1} \).
Therefore, by submultiplicativity the Lipschitz constant \( K \) for \( f \) is \( 2 \lambda_{\text{max}}((\mat{I} + \mat{L})^{-1})^2 \lambda_{\text{max}}(\mat{L})\).

Using the properties of eigenvalues we know that \(\lambda_{\text{max}}(1+\mat{X})^{-1}=\frac{1}{1+\lambda_{\text{min}}(\mat{X})} \); and since the minimum eigenvalue of a Laplacian matrix is 0, the maximum eigenvalue is 1.
Using the fact that the maximum eigenvalue of a graph Laplacian is bounded by the maximum degree in the graph, we get \( K= 2 \Delta(G)\).
\end{proof}

\begin{lemma}[Polarization plus Disagreement for undirected graphs (PD-Undir) ]
\label{lemma:pd-und}
Given a continuous differentiable function \( f(\mat{L}, \vec{s}) = \vec{s}^\top(\mat{I} + \mat{L})^{-1}\vec{s} \), where \(\mat{L}\) is a fixed symmetric Laplacian matrix and \(\vec{s} \in \mathbb{R}^n\), the function $f(\vec{s})$  is Lipschitz continuous with a Lipschitz constant equal to $2$.
\end{lemma}

\begin{proof}
From Eq.~\eqref{eq:gradient}, the gradient of \( f \) is \( \nabla_{\vec{s}} f(\mat{L}, \vec{s}) = 2(\mat{I} + \mat{L})^{-1}\vec{s} \).
The norm of this gradient ($\| 2(\mat{I} + \mat{L})^{-1}\vec{s} \|_2$) is bounded, by sub-multiplicativity, by the eigenvalues of \( (\mat{I} + \mat{L})^{-1} \).
Therefore, the Lipschitz constant \( K \) for \( f \) is \( 2 \lambda_{\text{max}}((\mat{I} + \mat{L})^{-1}) \).
Using the properties of eigenvalues we know that \(\lambda_{\text{max}}(1+\mat{A})^{-1}=\frac{1}{1+\lambda_{\text{min}}(\mat{A})} \); and since the minimum eigenvalue of a Laplacian matrix is 0, the maximum eigenvalue is 1.
Hence, \( K= 2 \lambda_{\text{max}}((\mat{I} + \mat{L})^{-1}) = 2 \).
\end{proof}

\section{Detailed Methodological Choices}
\subsection{Node Selection Strategies}
\label{app:node_sel}

Selecting the best $b$-element subset of nodes to query that minimizes the statistical error of reconstruction is NP-hard in general since it can be cased as discrete optimal design problems~\cite{pukelsheim2006, lorenzo2018sampling}, such as A-optimal design for minimizing the mean squared error.
To avoid computational bottlenecks, we devise three heuristic approaches that select the top-$b$ nodes based on the following centrality measures:\\
\noindent{\textit{Degree Centrality}}: This metric measures the total number of connections a node has, which includes both incoming and outgoing edges. Nodes with high degree centrality are considered influential due to their numerous direct connections.
It is defined as:
\begin{align*}
\text{Degree Centrality of node } u &= d_u^{\text{total}} = d_u^{\text{in}} + d_u^{\text{out}} \notag \\
&= \sum_{v \in V} A_{vu} + \sum_{v \in V} A_{uv}, 
\end{align*}
  where \( A \) is the adjacency matrix of the graph.

\noindent{\textit{Closeness Centrality}~\cite{bavelas1950communication}:} This metric measures how close a node is to all other nodes in the graph. It captures the average shortest path distance from a node to all other nodes, indicating the node's ability to quickly interact with all parts of the network. The measure is inversely proportional to the sum of the shortest path distances between a node and all others.
\begin{equation*}
  \text{Closeness Centrality of node } u = \frac{1}{\sum_{v \in V} p(u, v)}, 
\end{equation*}
  where \( p(u, v) \) is the shortest path distance between nodes \( u \) and \( v \).

\noindent{\textit{PageRank}~\cite{page1999pagerank}}: This metric measures the importance of a node based on the structure of incoming links, assigning higher importance to nodes with many incoming links from other important nodes. The PageRank of a node \( u \) is defined by the recursive formula
\[
u = PR(u) = \frac{1-d}{n} + d \sum_{v \in \text{In}(u)} \frac{PR(v)}{d_v}, 
\]
where \( d \) is a damping factor (set to 0.85), \( \text{In}(u) \) is the set of nodes that link to \( u \), and $d_v$ is the out-degree of $v$.

\section{Detailed Opinion Reconstruction Methods}
\spara{Label Propagation (LP).}
\label{app:pseudo}
Label Propagation (LP) is a semi-supervised learning algorithm originally designed for classification tasks, where labels are propagated through the network based on the labels of neighboring nodes~\cite{zhu2002learning}.
We extended it to handle continuous values, making it suitable for reconstructing opinions in a network. This approach leverages the fact that neighboring nodes tend to have similar opinions.

The LP strategy for continuous values initializes the known opinions for the selected nodes and iteratively updates the values of the remaining nodes based on the average values of their neighbors. This process continues until convergence or a maximum number of iterations is reached.

\spara{Graph Neural Networks (GNN).}
We use a GCN (Graph Convolutional Network)~\cite{kipf2016semi} to propagate the known opinions of the selected nodes through the network to reconstruct the unknown opinions.
The algorithm initializes the GCN with node features and trains it using the opinions of the selected nodes to define the squared error loss.
After training, the GCN predicts the values for all nodes in the graph.

In our implementation, we use the queried opinions as the node features if available, setting to zero if not available.
We also tested different structural features of the nodes, such as degree centrality, PageRank, and adjacency eigenvectors, but found no significant gains in accuracy, with only increased computational cost.

\spara{Graph Signal Processing (GSP).}
Graph Signal Processing (GSP) assumes that the graph signal (i.e., opinions in our case) can be described by a limited number \( |F| \) of frequencies and \( |X| \) nodes ($|X|=b$ in our problem) such that \( |F|, |X| \leq n \).
In GSP, frequencies correspond to eigenvectors of the Laplacian, which we denote with the column vectors in the matrix \( \mat{U} \in \mathbb{R}^{|V| \times |V|}\).
This signal can be written as 
\begin{equation}
\vec{s}_X = \mat{P}_X^\top \vec{s} = \mat{P}_X^\top \mat{U}_F \vec{\tilde{s}}_F,
\label{eq:sampled_signal}
\end{equation}
where \( \vec{s}_X \in \mathbb{R}^{|X|} \) is the observation vector over the vertex set \( X \subseteq V \), \( \mat{P}_X \in \mathbb{R}^{|V| \times |X|} \) is a sampling matrix whose columns are indicator functions for nodes in \( X \) and satisfies \( \mat{D}_X = \mat{P}_X \mat{P}_X^\top \), \( \mat{U}_F \in \mathbb{R}^{|F| \times n}\) is the matrix of the $F$ subset of Laplacian eigenvectors, and \( \vec{\tilde{s}}_F \) is a sparse representation of the opinions in the frequency domain.

The problem of recovering a graph signal that is limited in the frequency domain (hence called \textit{bandlimited}) from its samples is equivalent to the problem of properly selecting the sampling set \( X \) and then recovering \( \vec{s} \) from \( \vec{s}_X \) by inverting the system of equations in Eq.~\eqref{eq:sampled_signal}. Selecting an optimal sampling set \( X \) that maximizes a target cost function \( f \) is computationally challenging, as described in Section~\ref{sec:node_sel}.

In our implementation, we use two results from~\citet{lorenzo2018sampling}. First, the necessary and sufficient conditions for perfect recovery, which requires \( \| \mat{D}_{X^c} \mat{U}_F \|_2 < 1 \) and \( |X| \geq |F| \), where \( \mat{D}_{X^c} = \mat{I} - \mat{D}_X \) projects onto the complement vertex set \( X^c = V \setminus X \).
Second, we assume that the signal can be affected by noise, where the observation model is given by 

\begin{equation}
\vec{s}_X = \mat{P}_X^\top (\vec{s} + \vec{v}) = \mat{P}_X^\top \mat{U}_F \vec{\tilde{s}}_F + \mat{P}_X^\top \vec{v},
\end{equation}
where \( \vec{v} \) is a zero-mean noise vector with covariance matrix \( \mat{R}_v = \mathbb{E}\{\vec{v}\vec{v}^{\top}\} \). To design an interpolator in the presence of noise, we consider the best linear unbiased estimator (BLUE)~\cite{winer1971statistical}:

\begin{equation}
\hat{\vec{s}} = \mat{U}_F (\mat{U}_F^{\top} \mat{P}_X (\mat{P}_X^{\top} \mat{R}_v \mat{P}_X)^{-1} \mat{P}_X^{\top} \mat{U}_F)^{-1} \mat{U}_F^{\top} \mat{P}_S (\mat{P}_S^{\top} \mat{R}_v \mat{P}_X)^{-1} \vec{s}_X.
\label{eq:reconstr_with_error}
\end{equation}

\section{Pseudocodes}
\label{app:pseudo}
\subsection{Label Propagation (LP)}
All node values are initialized to zero (line 1), and the values of the sampled nodes are set to their corresponding real values (lines 2-4). For a fixed number of iterations, the algorithm updates the value of each node (except the sampled ones) to the average value of its neighbors (lines 5-15). Specifically, for each iteration, it creates a copy of the current node values (line 6) and then, for each node that is not sampled (lines 7-8), it calculates the average value of its neighbors if it has any (lines 9-11). The process continues until the maximum number of iterations is reached (lines 5 and 15). The resulting node values represent the recovered opinions for the entire graph (line 16).
Hence, the computational cost of Algorithm~\ref{alg:label_prop} is $O(\texttt{max\_iter} \times (|V| \times d_{\text{avg}}))$.

\subsection{Graph Neural Networks (GNN)}
In this method, all node features are initialized to zero (lines 2-3). The values of the sampled nodes are set to their corresponding real values (lines 4-5). A train mask is created to indicate which nodes have known values (lines 6-9). A GCN model is defined and initialized with the specified number of hidden channels. The model is trained for the specified number of epochs (lines 11-15), where it learns to predict node values by minimizing the MSE loss between the predicted and known values for the sampled nodes. After training, the model is used to predict the values for all nodes in the graph (line 16), and the predicted values are returned (line 17). 
Hence, the computational cost of Algorithm~\ref{alg:gnn_recovery} is $O(\texttt{epochs} \times |E| \times \texttt{hidden\_channels})$.

\begin{algorithm}[t!]
\caption{Label Propagation for Continuous Values}
\label{alg:label_prop}
\KwData{Graph $G = (V, E)$, sampled indices $S$, real values $\vec{s}_{\mathrm{true}}$, maximum iterations $max\_iter$}
\KwResult{Recovered node values $\vec{s}$}
$\vec{s}[i] \gets 0$ for all $i \in V$ \tcp*[f]{Initialize node values}\\
\For{each $i \in S$}{
    $\vec{s}[i] \gets \vec{s}_{\mathrm{true}}[i]$
}

\For{each iteration $t$ from $1$ to $max\_iter$}{
    $\vec{s}_{\text{new}} \gets \vec{s}$\\
    \For{each $i \in V$}{
        \If{$i \notin S$}{
            $N_i \gets \{j \in V \mid (i, j) \in E\}$ \tcp*[f]{Neighbors of $i$}
            \If{$N_i \neq \emptyset$}{
                $\vec{s}_{\text{new}}[i] \gets \frac{1}{|N_i|} \sum_{j \in N_i} \vec{s}[j]$
            }
        }
    }
    $\vec{s} \gets \vec{s}_{\text{new}}$
}
\Return $\vec{s}$
\end{algorithm}

\begin{algorithm}[t!]
\caption{Training of GNN-based Recovery}
\label{alg:gnn_recovery}
\KwData{Graph $G$, sampled indices $S$, real values $\vec{s}_{\mathrm{true}}$, $hidden\_channels$, number of $epochs$, learning rate $lr$}
\KwResult{Recovered node values $\vec{s}$}

\textbf{Input:} Graph $G$, node features $X$, targets $Y$ \\
$X \gets \mathbf{0}_{N \times 1}$ \tcp*[f]{Initialize node features}\\
\For{each $i \in S$}{
    $X[i] \gets \vec{s}_{\mathrm{true}}[i]$
}
$\mathrm{train\_mask} \gets \mathbf{0}_N$ \tcp*[f]{Initialize train mask}\\
\For{each $i \in S$}{
    $\mathrm{train\_mask}[i] \gets \text{True}$
}
Define and initialize the GCN model with $hidden\_channels$ \\
\For{epoch $i = 1$ to $epochs$}{
    $\hat{\vec{s}} \gets f_{\theta}(X, G)$ \tcp*[f]{Forward}\\
    $\mathrm{loss} \gets \frac{1}{|\text{train mask}|} \sum_{j \in \text{train mask}} (\hat{\vec{s}}[j] - \vec{s}_{\mathrm{true}}[j])^2$ \\
    \tcp{Backward pass and optimization step}
    $\theta \gets \mathrm{ADAM}(\theta, lr, \nabla_{\theta}\mathrm{loss}, \beta_1, \beta_2)$ \tcp*[f]{Backward}\\
}
$\vec{s} \gets f_{\theta}(X, G)$\tcp*[f]{Predicted values for all nodes}\\

\textbf{Output:} $\vec{s}$
\end{algorithm}

\section{Time Complexity}
The computational complexity of the overall algorithm depends on three main components: \textbf{node selection}, \textbf{innate opinion reconstruction}, and \textbf{opinion optimization}. Below, we provide the detailed time complexities for each step and the methods employed. 

\spara{Node Selection.}
The node selection step involves computing centrality metrics that determine the importance of nodes in the graph. The time complexities for the most common methods are as follows:

\begin{itemize}
    \item \textit{Closeness Centrality}: $\mathcal{O}(mn)$, where $m$ is the number of edges and $n$ is the number of nodes.
    \item \textit{Degree Centrality}: $\mathcal{O}(n \log n)$, representing the sorting of node degrees.
    \item \textit{PageRank}: $\mathcal{O}(m\tau)$, where $\tau$ is the number of iterations of the power method used to compute PageRank.
\end{itemize}

\spara{Innate Opinion Reconstruction.}
The reconstruction of innate opinions from observed graph data can be performed using various methods, each with distinct computational costs:

\begin{itemize}
    \item \textit{Graph Neural Networks (GNN)}: $\mathcal{O}(\text{epochs} \cdot m \cdot \text{hidden}_{\text{channels}})$, where \textit{epochs} is the number of training iterations, $m$ is the edge count, and $\text{hidden}_{\text{channels}}$ refers to the size of hidden layers in the GNN (see Appendix B.1 for details).
    
    \item \textit{Graph Signal Processing (GSP)}: $\mathcal{O}(n^3)$, due to the matrix inversion and eigendecomposition required to reconstruct signals.
    
    \item \textit{Label Propagation (LP)}: $\mathcal{O}(\max\{\text{iter} \cdot n \cdot d_{\text{avg}}\})$, where \textit{iter} is the number of iterations, $n$ is the node count, and $d_{\text{avg}}$ is the average node degree.
\end{itemize}

\spara{Opinion Optimization.}
The opinion optimization step focuses on minimizing both \textit{polarization} and \textit{disagreement} in the graph. The time complexity depends heavily on the gradient-based methods and the stopping conditions chosen. In the directed case, the time complexity is given by:

\begin{equation}
\mathcal{O}(T(m + n)\tau),
\end{equation}
where $T$ is the number of iterations of the BiConjugate Gradient Stabilized (BiCGStab) solver in each optimization step, $m$ is the number of edges, $n$ is the number of nodes, and $\tau$ is the number of iterations of the gradient-based algorithm.

\section{Running Time Analysis}

We report the running time of our method when applied to real-world networks. The objective function used in the experiments is \textbf{PD-Dir}, and the node selection strategy is based on degree centrality with $b = 0.2n$.

Table~\ref{tab:running-times} summarizes the running times of three different methods—\textbf{GNN} (Graph Neural Networks), \textbf{GSP} (Graph Signal Processing), and \textbf{LP} (Label Propagation)—applied to three real-world networks: \textit{brexit}, \textit{referendum}, and \textit{vaxNoVax}.

\begin{table}[h]
    \centering
    \caption{Running times of methods on real-world networks.}
    \label{tab:running-times}
    \begin{tabular}{l l r}
        \hline
        \textbf{Model} & \textbf{Network} & \textbf{Time (sec)} \\ 
        \hline
        GNN & brexit       & 231.64 \\
        GNN & referendum   & 26.24  \\
        GNN & vaxNoVax     & 685.95 \\
        GSP & brexit       & 768.04 \\
        GSP & referendum   & 40.51  \\
        GSP & vaxNoVax     & 680.71 \\
        LP  & brexit       & 1431.46 \\
        LP  & referendum   & 174.62 \\
        LP  & vaxNoVax     & 4213.62 \\
        \hline
    \end{tabular}
\end{table}

The results highlight that the GNN-based approach is the most scalable method, demonstrating superior efficiency for large-scale networks such as \textit{vaxNoVax} with 1.6M edges. GSP shows acceptable scalability but becomes computationally intensive on larger graphs. In contrast, LP struggles with scalability.

\section{Experimental Setup}
\label{sec:setup-appendix}

\spara{Graph Instances.}
We consider 11 real-world directed graphs and 5 undirected graphs related to social media/networks, ranging from 30 to 12,000 nodes and up to 1,600,000 edges. 
Table~\ref{tab:datasets} summarizes the statistics of the datasets.
The first three datasets contain the follow network on $\mathbb{X}$ and are derived from~\citet{minici2022cascade}. 
The other networks are related to social activities and obtained from KONECT\footnote{\url{http://konect.cc/}}~\cite{kunegis2013konect}. 

We preprocess the graphs by removing disconnected nodes with zero outdegree, self-loops, and multiple edges. 
We ensure that all weights, if available, are positive. In the case of directed networks,  we row-normalize the adjacency matrix as in \citet{cinus2023rebalancing}.
A directed edge $(u,v)$ indicates that $u$ has visibility of $v$'s contents, and hence $u$ can be influenced by $v$. 

\spara{Opinions Instances.}
The \texttt{Referendum}, \texttt{Brexit}, and \texttt{VaxNoVax} datasets contain exposed opinions for each node, which are the average stances of tweets retweeted by the users; each tweet's stance was estimated with a supervised text-based classifier~\cite{minici2022cascade}.
For graphs where such real proxies of opinions are not available, we generate opinions by sampling from either a uniform distribution between $[-0.5, +0.5]$ or from Gaussian distributions. 
To induce structural polarization, we use the Kernighan-Lin bisection method~\cite{kernighan1970efficient} to identify two communities in the graph.
We then assign to each community average opinions symmetrically distant around 0.
Node opinions are subsequently sampled from the Gaussian distribution centered around the average opinion assigned to their community.
We preprocess opinions by inserting polarization in the statistical distribution with a parameter \( p \) using the function $f(x | p) =  \left| x \right|^{\frac{1}{p}}$, using $p=3$.
We assume that the available opinions are those at equilibrium, hence we invert the FJ model dynamics to recover the innate opinions from the exposed one.
Then we mean-centered the distribution according to~\citet{musco2018minimizing} so that the average innate opinion is zero and we set its standard deviation is 1 (Table~\ref{tab:datasets} shows the samples of these distributions).

\spara{Parameters.}
In the objective functions for directed graphs, to obtain fast convergence to a local minimum, the directed problems are solved using ADAM with $\beta_1 = 0.9$, $\beta_2 = 0.999$, and learning rate $\eta = 0.2$. We use early stopping when the relative change of the objective between different steps is less than $20\%$.
For the problem with \textsc{PD-Undir}, to reach the global optimum, we used the Splitting Conic Solver (SCS)~\cite{ocpb:16,odonoghue:21} of CVXPY with 2,500 maximum number of iterations and a tolerance of $10^{-3}$.

In opinion reconstruction, we set the querying budget $b$ to 20\% of nodes which enables fair comparison across all experiments, methods, and datasets. 
We also tested the effect of the number of budget $b$ in reconstruction and optimization (See Figures ~\ref{fig:sensors-numb} and ~\ref{fig:sensors-numb-gsignal}).
In the GSP strategy, we set the number of frequencies equal to 15\% of nodes to satisfy the $b \geq |F|$. 
The dependency with respect to the number of frequencies and sampled nodes is reported in Figure~\ref{fig:sensors-numb-gsignal}.
We also symmetrize the adjacency matrix to use a unique standard GSP method for undirected graphs.
In the GNN strategy, we used GCN model~\cite{kipf2016semi} with 16 hidden channels, 200 epochs, and a learning rate of 0.01.
In the LP strategy, we used 200 iterations. The above parameter settings were empirically evaluated to obtain a tradeoff between convergence and running time.

\section{Additional Experiments}
\subsection{Performance in Directed graphs with uniform opinion distribution}
Results are presented in Table~\ref{tab:directed-opt-uniform} are consistent with those in the main text. 
Here we consider the 8 real-world directed networks in Table~\ref{tab:datasets} with a uniform opinion distribution.

In general, similarly to the polarized opinions configuration, our pipeline yields multiplicative errors usually below 2 compared to directly optimizing over error-free opinions. 
The LP reconstruction methodology consistently achieves multiplicative errors below 2; except for the ``oz'' network, which is consistently difficult to optimize for all methods. 
In ``ciaodvd-trust" now all methods achieve comparable results, meaning that the main difficulty for the methods was the presence of structural polarization.

\begin{table}[h!]
\caption{Average Multiplicative Error for the three objectives (\textsc{P-Dir}, \textsc{D-Dir}, \textsc{PD-Dir}) for 8 real-world directed graphs with different sizes ($n$).
Opinions are uniformly distributed and reconstructed with $b=0.20|V|$ nodes selected based on their degree.} \label{tab:directed-opt-uniform}
\let\center\empty
\let\endcenter\relax
\centering
\begin{tabular}{cc|ccc}
\toprule
   &                       & \multicolumn{3}{l}{Multiplicative Error} \\
   &        Rec Method     &       GNN &            GSP &          LP \\
Objective & Network        &           &                 &            \\
\midrule
\textsc{P-Dir} & highschool         &    \(2.02 \pm 0.25\) &  \(1.97 \pm 0.24\) &  \(\mathbf{1.90 \pm 0.20}\) \\
               & wiki talk         &    \(1.27 \pm 0.23\) &  \(\mathbf{1.21 \pm 0.13}\) &  \(1.33 \pm 0.24\) \\
               & innovation        &    \(1.89 \pm 0.14\) &  \(1.96 \pm 0.20\) &  \(\mathbf{1.81 \pm 0.13}\) \\
               & oz                &   \(1.97 \pm 0.10\) &  \(1.93 \pm 0.08\) &  \(\mathbf{1.88 \pm 0.09}\) \\
               & filmtrust         &    \(1.33 \pm 0.05\) &  \(1.35 \pm 0.05\) &  \(\mathbf{1.27 \pm 0.04}\) \\
               & dnc-temporal &   \(1.24 \pm 0.05\) &  \(\mathbf{1.21 \pm 0.05}\) &  \(1.28 \pm 0.11\) \\
               & ciaodvd-trust     &   \(1.76 \pm 0.05\) &  \(1.86 \pm 0.07\) &  \(\mathbf{1.64 \pm 0.04}\) \\
               & health     &   \(1.82 \pm 0.03\) &  \(1.83 \pm 0.04\) &  \(\mathbf{1.72 \pm 0.03}\) \\
\midrule
\textsc{D-Dir} & highschool         &   \(1.67 \pm 0.13\) &  \(1.67 \pm 0.10\) &  \(\mathbf{1.62 \pm 0.11}\) \\
               & wiki talk         &   \(1.20 \pm 0.16\) &  \(\mathbf{1.18 \pm 0.09}\) &  \(1.28 \pm 0.16\) \\
               & innovation        &   \(1.82 \pm 0.14\) &  \(1.86 \pm 0.11\) &  \(\mathbf{1.67 \pm 0.10}\) \\
               & oz                &   \(2.61 \pm 0.12\) &  \(2.52 \pm 0.13\) &  \(\mathbf{2.46 \pm 0.11}\) \\
               & filmtrust         &   \(1.24 \pm 0.03\) &  \(1.24 \pm 0.03\) &  \(\mathbf{1.23 \pm 0.03}\) \\
               & dnc-temporal &   \(1.31 \pm 0.05\) &  \(\mathbf{1.23 \pm 0.06}\) &  \(1.40 \pm 0.08\) \\
               & ciaodvd-trust     &   \(2.01 \pm 0.06\) &  \(1.99 \pm 0.05\) &  \(\mathbf{1.75 \pm 0.03}\) \\
               & health            &   \(1.54 \pm 0.03\) &  \(1.54 \pm 0.02\) &  \(\mathbf{1.45 \pm 0.02}\) \\
\midrule

\textsc{PD-Dir} & highschool        &   \(1.35 \pm 0.08\) &  \(1.31 \pm 0.08\) &  \(\mathbf{1.27 \pm 0.08}\) \\
                & wiki talk         &   \(1.17 \pm 0.18\) &  \(\mathbf{1.13 \pm 0.09}\) &  \(1.23 \pm 0.19\) \\
                & innovation        &   \(1.30 \pm 0.06\) &  \(1.31 \pm 0.07\) &  \(\mathbf{1.24 \pm 0.05}\) \\
                & oz                &   \(1.37 \pm 0.05\) &  \(1.35 \pm 0.03\) &  \(\mathbf{1.34 \pm 0.05}\) \\
                & filmtrust         &   \(1.14 \pm 0.03\) &  \(1.13 \pm 0.02\) &  \(\mathbf{1.08 \pm 0.02}\) \\
                & dnc-temporal &   \(1.11 \pm 0.03\) &  \(\mathbf{1.09 \pm 0.03}\) &  \(\mathbf{1.09 \pm 0.07}\) \\
                & ciaodvd-trust     &   \(1.30 \pm 0.02\) &  \(1.27 \pm 0.02\) &  \(\mathbf{1.19 \pm 0.02}\) \\
                & health            &   \(1.29 \pm 0.01\) &  \(1.28 \pm 0.01\) &  \(\mathbf{1.24 \pm 0.01}\) \\
\bottomrule
\end{tabular}

\end{table}

\subsection{Performance in Undirected graphs with uniform opinion distribution}
Results are presented in Table~\ref{tab:undirected-opt-unif}. We consider the 5 real-world undirected networks in Table~\ref{tab:datasets} with uniform opinion distribution. 

Reducing polarization and disagreement in a non-polarized configuration is more difficult if solved with reconstructed opinions compared to polarized configurations.
In fact, multiplicative errors are higher compared to the ones in Table~\ref{tab:undirected-opt} in the main text.
Again, errors show greater variability across different graphs in undirected settings compared to directed ones. This is because the global optimum corresponds to numerically low levels of polarization and disagreement. As a result, the denominator in the multiplicative error calculation is often smaller than 1, amplifying the observed errors.
Given that opinions are uniform and so they reflect no structure, bounds are both sensitive to the difficulty of opinion reconstruction (which exploits the graph structure) and the numerical value of the global minimum of the objective. 
As a result, the bounds are larger than the actual error by one order of magnitude, indicating that, in practice, the problem is less challenging than theoretically predicted, and a tighter bound likely exists at this graph size scale.

\begin{table}[h!]
\caption{Average Multiplicative error in \textsc{PD-Undir} minimization in undirected graphs with uniform opinion distribution.} 
\label{tab:undirected-opt-unif}
\let\center\empty
\let\endcenter\relax
\centering
\begin{tabular}{c|ccc}
\toprule
           {} & \multicolumn{3}{l}{Multiplicative Error} \\
Rec Method    &                     GNN       &                          GSP  &                       LP     \\
 Network      &                               &                               &                              \\
\midrule
 zachary     &       \(3.70 \pm 0.78 \:\: (10)\) &      \(\mathbf{2.74 \pm 0.59} \:\: (11)\) &     \(3.67 \pm 0.92 \:\: (11)\) \\
 beach       &      \(\mathbf{6.67 \pm 4.12} \:\: (138)\) &    \(10.93 \pm 5.12 \:\: (166)\) &    \(7.37 \pm 7.12 \:\: (104)\) \\
 train       &       \(\mathbf{3.64 \pm 1.18} \:\: (11)\) &      \(3.84 \pm 1.07 \:\: (12)\) &     \(4.24 \pm 1.17 \:\: (11)\) \\
 mit         &    \(\mathbf{0.99 \pm 0.03} \:\: (10763)\) &   \(1.00 \pm 0.01 \:\: (12132)\) &   \(1.00 \pm 0.02 \:\: (7432)\) \\
 football    &       \(6.31 \pm 1.12 \:\: (14)\) &      \(\mathbf{6.25 \pm 1.20} \:\: (15)\) &     \(7.00 \pm 1.33 \:\: (13)\) \\
\bottomrule
\end{tabular}

\end{table}

\subsection{Performance of Node Selection Strategies in Directed graphs}
Also in directed graphs with synthetic opinions, \textsf{degree} and \textsf{PageRank} are the best strategies for selecting nodes. 
Compared to real datasets, the increment with respect to random selection is smaller.
Furthermore, in the ``wiki talk'' network random selection is comparable or superior to other methods.
We hypothesize that this is due to the unique structural characteristics of the graph. Notably, the ``wiki talk ht'' and ``dnc-temporalGraph'' networks exhibit the lowest degree assortativity coefficient (-0.4), indicating that highly connected nodes tend to connect with low-degree peripheral ones. In such cases, labels propagating from hubs (selected by degree-based methods) are likely to remain confined within localized areas, which can lead to suboptimal performance.
Additionally, we highlight that the Random Strategy is far from an arbitrary or ineffective approach. Selecting nodes uniformly at random often results in coverage of diverse regions of the network. Given the strong homophily typically observed in real-world innate opinions, this strategy can provide a reasonable basis for querying nodes to infer their innate opinions effectively.

Finally, results are consistent across the two different opinion distributions: polarized~\ref{tab:sampling-strategies-synth-gauss} and uniform~\ref{tab:sampling-strategies-synth-unif}.
``ciaodvd-trust'' results the most difficult network with polarized opinions, with multiplicative errors above 2 while reducing \textsc{P-Dir}.
On the contrary, the non-polarized configuration in the ``oz'' dataset has the highest multiplicative errors in disagreement minimization.

\begin{table*}
\caption{Multiplicative Error for the three objectives (\textsc{P-Dir}, \textsc{D-Dir}, \textsc{PD-Dir}) for different node selection strategies in directed networks with polarized opinions.} 
\label{tab:sampling-strategies-synth-gauss}
\let\center\empty
\let\endcenter\relax
\centering
\begin{tabular}{cc|cccc}
\toprule
          &    {}         & \multicolumn{4}{l}{Multiplicative Error} \\
          &    Sel Method & Closeness centrality &             Degree &           PageRank &             Random \\
Objective &    Network    &                      &                     &                    &                   \\
\midrule
\textsc{P-Dir} & highschool         &       \(2.01 \pm 0.42\) &  \(\mathbf{1.68 \pm 0.28}\) &  \(1.71 \pm 0.26\) &  \(1.95 \pm 0.33\) \\
               & wiki talk         &       \(\mathbf{1.31 \pm 0.24}\) &  \(1.34 \pm 0.27\) &  \(1.35 \pm 0.27\) &  \(1.32 \pm 0.21\) \\
               & innovation        &       \(1.77 \pm 0.20\) &  \(1.80 \pm 0.21\) &  \(\mathbf{1.73 \pm 0.18}\) &  \(2.15 \pm 0.32\) \\
               & oz                &       \(1.80 \pm 0.20\) &  \(1.78 \pm 0.17\) &  \(\mathbf{1.77 \pm 0.19}\) &  \(1.82 \pm 0.17\) \\
               & filmtrust-trust   &       \(1.32 \pm 0.07\) &  \(1.27 \pm 0.06\) &  \(\mathbf{1.25 \pm 0.06}\) &  \(1.40 \pm 0.10\) \\
               & dnc-temporal      &       \(1.28 \pm 0.16\) &  \(1.25 \pm 0.14\) &  \(\mathbf{1.24 \pm 0.13}\) &  \(1.33 \pm 0.10\) \\
               & ciaodvd-trust     &       \(2.53 \pm 0.14\) &  \(2.54 \pm 0.14\) &  \(2.43 \pm 0.15\) &  \(\mathbf{2.01 \pm 0.10}\) \\
               & health            &       \(1.60 \pm 0.08\) &  \(\mathbf{1.57 \pm 0.06}\) &  \(\mathbf{1.57 \pm 0.07}\) &  \(1.68 \pm 0.08\) \\
\midrule
\textsc{D-Dir} & highschool         &       \(1.51 \pm 0.23\) &  \(\mathbf{1.42 \pm 0.15}\) &  \(1.48 \pm 0.18\) &  \(1.51 \pm 0.21\) \\
               & wiki talk         &       \(1.36 \pm 0.20\) &  \(1.34 \pm 0.17\) &  \(1.35 \pm 0.21\) &  \(\mathbf{1.28 \pm 0.16}\) \\
               & innovation        &       \(1.52 \pm 0.20\) &  \(\mathbf{1.51 \pm 0.15}\) &  \(1.52 \pm 0.19\) &  \(1.71 \pm 0.21\) \\
               & oz                &       \(1.47 \pm 0.16\) &  \(1.47 \pm 0.14\) &  \(\mathbf{1.45 \pm 0.14}\) &  \(1.57 \pm 0.13\) \\
               & filmtrust         &       \(\mathbf{1.20 \pm 0.04}\) &  \(1.22 \pm 0.05\) &  \(1.22 \pm 0.04\) &  \(1.25 \pm 0.07\) \\
               & dnc-temporal      &       \(1.33 \pm 0.10\) &  \(1.35 \pm 0.11\) &  \(1.35 \pm 0.11\) &  \(\mathbf{1.29 \pm 0.09}\) \\
               & ciaodvd-trust     &       \(1.51 \pm 0.03\) &  \(1.52 \pm 0.03\) &  \(1.50 \pm 0.03\) &  \(\mathbf{1.48 \pm 0.04}\) \\
               & health            &       \(1.37 \pm 0.05\) &  \(\mathbf{1.36 \pm 0.04}\) &  \(\mathbf{1.36 \pm 0.05}\) &  \(1.40 \pm 0.05\) \\
\midrule
\textsc{PD-Dir} & highschool        &       \(1.47 \pm 0.19\) &  \(\mathbf{1.32 \pm 0.09}\) &  \(1.36 \pm 0.12\) &  \(1.46 \pm 0.13\) \\
                & wiki talk         &       \(\mathbf{1.26 \pm 0.22}\) &  \(1.28 \pm 0.23\) &  \(1.29 \pm 0.23\) &  \(\mathbf{1.26 \pm 0.19}\) \\
                & innovation        &       \(1.36 \pm 0.10\) &  \(1.35 \pm 0.10\) &  \(\mathbf{1.34 \pm 0.11}\) &  \(1.53 \pm 0.15\) \\
                & oz                &       \(\mathbf{1.36 \pm 0.10}\) &  \(1.40 \pm 0.09\) &  \(1.38 \pm 0.11\) &  \(1.41 \pm 0.11\) \\
                & filmtrust         &       \(1.18 \pm 0.04\) &  \(1.15 \pm 0.03\) &  \(\mathbf{1.13 \pm 0.04}\) &  \(1.23 \pm 0.06\) \\
                & dnc-temporal      &       \(1.20 \pm 0.13\) &  \(\mathbf{1.16 \pm 0.12}\) &  \(\mathbf{1.16 \pm 0.11}\) &  \(1.22 \pm 0.10\) \\
                & ciaodvd-trust     &       \(1.55 \pm 0.05\) &  \(1.56 \pm 0.05\) &  \(1.52 \pm 0.05\) &  \(\mathbf{1.30 \pm 0.04}\) \\
                & health            &       \(\mathbf{1.29 \pm 0.03}\) &  \(\mathbf{1.29 \pm 0.02}\) &  \(\mathbf{1.29 \pm 0.03}\) &  \(1.34 \pm 0.04\) \\
\bottomrule
\end{tabular}

\end{table*}

\begin{table*}
\caption{Multiplicative Error for the three objectives (\textsc{P-Dir}, \textsc{D-Dir}, \textsc{PD-Dir}) for different node selection strategies in directed networks with uniform opinion distribution.} 
\label{tab:sampling-strategies-synth-unif}
\let\center\empty
\let\endcenter\relax
\centering
\begin{tabular}{cc|cccc}
\toprule
          &   {}         &    \multicolumn{4}{l}{Multiplicative Error} \\
          &   Sel Method & Closeness centrality &             Degree &           PageRank &             Random \\
Objective & Network      &                      &                    &                    &                    \\
\midrule
\textsc{P-Dir} & highschool           &    \(1.89 \pm 0.26\) &  \(1.90 \pm 0.20\) &  \(\mathbf{1.84 \pm 0.21}\) &  \(1.91 \pm 0.21\) \\
               & wiki talk           &    \(1.34 \pm 0.21\) &  \(1.33 \pm 0.24\) &  \(1.36 \pm 0.25\) &  \(\mathbf{1.31 \pm 0.21}\) \\
               & innovation          &    \(\mathbf{1.80 \pm 0.13}\) &  \(1.81 \pm 0.13\) &  \(\mathbf{1.80 \pm 0.13}\) &  \(1.86 \pm 0.15\) \\
               & oz                  &    \(\mathbf{1.84 \pm 0.09}\) &  \(1.88 \pm 0.09\) &  \(1.86 \pm 0.08\) &  \(1.93 \pm 0.09\) \\
               & filmtrust           &    \(1.30 \pm 0.04\) &  \(\mathbf{1.27 \pm 0.04}\) &  \(1.31 \pm 0.05\) &  \(1.37 \pm 0.05\) \\
               & dnc-temporal        &    \(1.29 \pm 0.12\) &  \(1.28 \pm 0.11\) &  \(1.29 \pm 0.11\) &  \(\mathbf{1.27 \pm 0.07}\) \\
               & ciaodvd-trust       &    \(\mathbf{1.63 \pm 0.03}\) &  \(1.64 \pm 0.04\) &  \(1.64 \pm 0.04\) &  \(1.70 \pm 0.04\) \\
               & health              &    \(\mathbf{1.70 \pm 0.03}\) &  \(1.72 \pm 0.03\) &  \(1.71 \pm 0.03\) &  \(1.79 \pm 0.04\) \\
\midrule
\textsc{D-Dir} & highschool           &    \(1.61 \pm 0.11\) &  \(1.62 \pm 0.11\) &  \(\mathbf{1.58 \pm 0.10}\) &  \(1.66 \pm 0.10\) \\
               & wiki talk           &    \(1.31 \pm 0.14\) &  \(1.28 \pm 0.16\) &  \(1.28 \pm 0.13\) &  \(\mathbf{1.22 \pm 0.14}\) \\
               & innovation          &    \(1.68 \pm 0.11\) &  \(\mathbf{1.67 \pm 0.10}\) &  \(1.69 \pm 0.10\) &  \(1.83 \pm 0.13\) \\
               & oz                  &    \(\mathbf{2.44 \pm 0.11}\) &  \(2.46 \pm 0.11\) &  \(2.45 \pm 0.13\) &  \(2.56 \pm 0.11\) \\
               & filmtrust           &    \(1.24 \pm 0.02\) &  \(\mathbf{1.23 \pm 0.03}\) &  \(1.25 \pm 0.03\) &  \(1.27 \pm 0.03\) \\
               & dnc-temporal        &    \(1.37 \pm 0.08\) &  \(1.40 \pm 0.08\) &  \(1.41 \pm 0.08\) &  \(\mathbf{1.36 \pm 0.05}\) \\
               & ciaodvd-trust       &    \(\mathbf{1.75 \pm 0.03}\) &  \(\mathbf{1.75 \pm 0.03}\) &  \(\mathbf{1.75 \pm 0.03}\) &  \(2.01 \pm 0.06\) \\
               & health              &    \(\mathbf{1.45 \pm 0.02}\) &  \(\mathbf{1.45 \pm 0.02}\) &  \(\mathbf{1.45 \pm 0.02}\) &  \(1.54 \pm 0.02\) \\
\midrule
\textsc{PD-Dir} & highschool          &    \(\mathbf{1.26 \pm 0.10}\) &  \(1.27 \pm 0.08\) &  \(1.27 \pm 0.08\) &  \(1.32 \pm 0.09\) \\
                & wiki talk           &    \(1.23 \pm 0.20\) &  \(1.23 \pm 0.19\) &  \(1.26 \pm 0.24\) &  \(\mathbf{1.22 \pm 0.19}\) \\
                & innovation          &    \(1.25 \pm 0.06\) &  \(\mathbf{1.24 \pm 0.05}\) &  \(1.25 \pm 0.06\) &  \(1.32 \pm 0.10\) \\
                & oz                  &    \(\mathbf{1.31 \pm 0.04}\) &  \(1.34 \pm 0.05\) &  \(1.34 \pm 0.05\) &  \(1.36 \pm 0.05\) \\
                & filmtrust           &    \(1.11 \pm 0.02\) &  \(\mathbf{1.08 \pm 0.02}\) &  \(1.10 \pm 0.02\) &  \(1.16 \pm 0.03\) \\
                & dnc-temporal        &    \(1.10 \pm 0.07\) &  \(\mathbf{1.09 \pm 0.07}\) &  \(\mathbf{1.09 \pm 0.07}\) &  \(1.16 \pm 0.05\) \\
                & ciaodvd-trust       &    \(\mathbf{1.19 \pm 0.01}\) &  \(\mathbf{1.19 \pm 0.02}\) &  \(\mathbf{1.19 \pm 0.01}\) &  \(1.29 \pm 0.02\) \\
                & health              &    \(\mathbf{1.23 \pm 0.02}\) &  \(1.24 \pm 0.01\) &  \(1.24 \pm 0.01\) &  \(1.29 \pm 0.01\) \\
\bottomrule
\end{tabular}

\end{table*}

\begin{figure}[h!]
\centering
\includegraphics[width=1.\columnwidth]{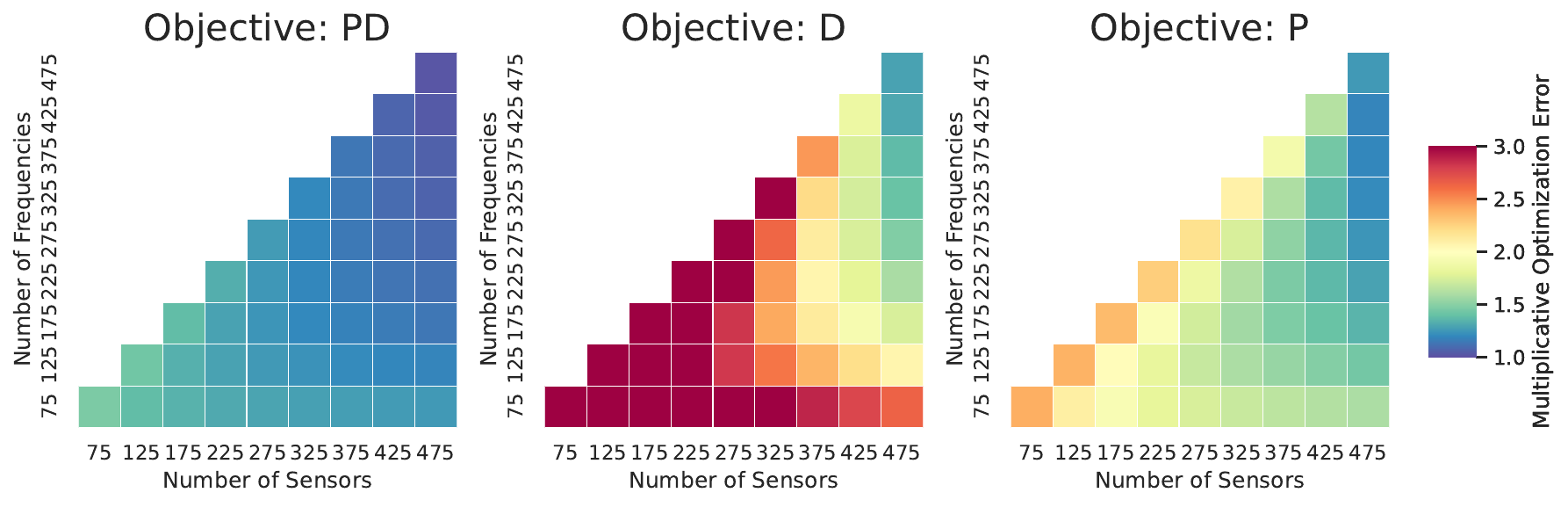}
\caption{Average multiplicative error vs the number of selected nodes vs the number of frequencies. We used an Erdos Renyi graph model with $|V|=500, p=0.25$, and polarized opinions.}
\label{fig:sensors-numb-gsignal}
\end{figure}

\subsection{Sensitivity Analysis in Graph Signal Processing}
In Figure~\ref{fig:sensors-numb-gsignal} we quantify the multiplicative error induced by the selected nodes size and frequencies using the Graph Signal Processing reconstruction strategy. 
Recalling that the number of frequencies must be lower or equal to the number of selected nodes ~\cite{lorenzo2018sampling}, we observe that multiplicative errors monotonically decrease along both dimensions as expected. Furthermore, the disagreement objective is the most difficult to optimize.

\subsection{Additional Baseline}
To further evaluate our approach, we tested an additional baseline where reconstructed innate opinions were assigned values drawn uniformly at random from the range of selected seed opinions.

We compared the performance of LP Opinion Reconstruction (LP) against this Random baseline under the condition where innate opinions were sampled from a Gaussian distribution, with the objective of minimizing \textsc{P-Dir}. The results indicate that the Random baseline consistently underperforms compared to LP Opinion Reconstruction.

Results are reported in Table~\ref{tab:random-baseline}.

\begin{table}
\caption{Multiplicative Error for the \textsc{P-Dir} objective for degree node selection strategy in directed networks with gaussian opinion distribution.} 
\label{tab:random-baseline}
\let\center\empty
\let\endcenter\relax
\centering
\begin{tabular}{cc|cccc}
\toprule
 {}         &    \multicolumn{2}{l}{Multiplicative Error} \\
 Sel Method & LP &             Random  \\
    Network &    &                     \\
\midrule
highschool   & \(\mathbf{1.90 \pm 0.20}\)         & \(2.26 \pm 0.34\) \\
wiki talk    & \(\mathbf{1.33 \pm 0.24}\)         & \(1.36 \pm 0.15\) \\
innovation   & \(\mathbf{1.81 \pm 0.13}\)         & \(2.46 \pm 0.22\) \\
oz           & \(\mathbf{1.88 \pm 0.09}\)         & \(2.92 \pm 0.14\) \\
filmtrust    & \(\mathbf{1.27 \pm 0.04}\)         & \(1.65 \pm 0.09\) \\
temporal     & \(\mathbf{1.28 \pm 0.11}\)         & \(1.55 \pm 0.10\) \\
ciaodvd      & \(\mathbf{1.64 \pm 0.04}\)         & \(2.73 \pm 0.10\) \\
health       & \(\mathbf{1.72 \pm 0.03}\)         & \(2.08 \pm 0.07\) \\
\bottomrule
\end{tabular}
\end{table}

\end{document}